\documentclass[preprint]{aastex62}
\usepackage{makecell}
\usepackage{gensymb}
\usepackage{multirow}
\usepackage{footnote}
\usepackage{url}
\newcommand{\RN}[1]{%
  \textup{\uppercase\expandafter{\romannumeral#1}}%
}
\newcommand{\hours}{\ensuremath{^\mathrm{h}}}
\newcommand{\minutes}{\ensuremath{^\mathrm{m}}}
\newcommand{\seconds}{\ensuremath{^\mathrm{s}}}

\shorttitle{Physical and Chemical Conditions of HL Tau}
\shortauthors{Wu et al.}
\begin{document}
\title{Physical and Chemical Conditions\\ of the Protostellar Envelope and the Protoplanetary Disk in HL Tau}
\correspondingauthor{Chun-Ju Wu}
\email{cjwu@asiaa.sinica.edu.tw}

\author{Chun-Ju Wu}
\affiliation{Department of Physics, National Taiwan University, No. 1, Sec. 4, Roosevelt Road, Taipei 106, Taiwan}
\affiliation{Academia Sinica Institute of Astronomy and Astrophysics, P.O. Box 23-141, Taipei 10617, Taiwan}

\author{Naomi Hirano}
\affiliation{Academia Sinica Institute of Astronomy and Astrophysics, P.O. Box 23-141, Taipei 10617, Taiwan}

\author{Shigehisa Takakuwa}
\affiliation{Academia Sinica Institute of Astronomy and Astrophysics, P.O. Box 23-141, Taipei 10617, Taiwan}
\affiliation{Department of Physics and Astronomy, Graduate School of Science and Engineering, Kagoshima University, 1-21-35 Korimoto, Kagoshima, Kagoshima 890-0065, Japan}

\author{Hsi-Wei Yen}
\affiliation{European Southern Observatory (ESO), Karl-Schwarzschild-Str. 2, D-85748 Garching, Germany}

\author{Yusuke Aso}
\affiliation{Academia Sinica Institute of Astronomy and Astrophysics, P.O. Box 23-141, Taipei 10617, Taiwan}

\begin{abstract}
We report our SMA observations of the Class \RN{1}-\RN{2} protostar HL Tau in the $^{13}$CO (2--1), C$^{18}$O (2--1), SO(5$_6$--4$_5$), and the 1.3 mm dust-continuum emission and our analyses of the ALMA long baseline data of the HCO$^{+}$ (1--0) emission. 
The 1.3 mm continuum emission observed with the SMA shows compact ($\sim$0$\farcs$8 $\times$ 0$\farcs$5) and extended ($\sim$6$\farcs$5 $\times$ 4$\farcs$3) components, tracing the protoplanetary disk and the protostellar envelope, respectively. 
The $^{13}$CO, C$^{18}$O, and HCO$^+$ show a compact ($\sim$ 200 AU) component at velocities higher than 3 km s$^{-1}$ from the systemic velocity and an extended ($\sim$ 1000 AU) component at the lower velocities.
The high-velocity component traces the Keplerian rotating disk, and the low-velocity component traces the infalling envelope.
The HCO$^+$ high-velocity component is fitted with a Keplerian disk model with a central stellar mass of 1.4 $M_{\odot}$. 
The radial intensity profiles of the $^{13}$CO and C$^{18}$O along the disk major axis are fitted with a disk+envelope model,  
and the gas masses of the disk and envelope are estimated to be $2\mbox{--}40\times10^{-4}$ $M_{\odot}$ and $2.9\times10^{-3}$ $M_{\odot}$, respectively.
The disk dust mass has been estimated to be $1\mbox{--}3 \times 10^{-3}$ $M_{\odot}$ in the literature.
Thus, our estimated disk gas mass suggests that the gas-to-dust mass ratio in the disk is $<$10, a factor of ten lower than the estimated ratio in the envelope. 
We discuss the possible gas depletion or CO depletion in the planet-forming candidate HL Tau  in the context of disk and envelope evolution.

\end{abstract}


\section{Introduction}
HL Tau is a Class \RN{1}-\RN{2} protostar ($L_{bol} >$ 3 $L_{\odot}$; $T_{bol} \sim$ 576 K; \citet{sta95,whi04}) located in the L1551 region at a distance of 140 pc \citep{reb04}. The source is surrounded by an infalling envelope with a size of 3000 au \citep{hay93,cab96,clo97,wel00}, and is associated with optical and infrared jets as well as a bipolar molecular outflow along the northeast to southwest direction \citep{mun90,mon96,tak07,hay09,lum14}. HL Tau is thus considered to be still in the active mass accretion phase. The circumstellar disk with a radius of $\sim$100 au around HL Tau has been identified and subjected to the detailed studies of the internal physical structures \citep{mun96,wil96,car09,kwo11,car16}.

The discovery of seven ringlike gaps in the millimeter and submillimeter dust-continuum emission in the circumstellar disk around HL Tau with the Atacama Large Millimeter/Submillimeter Array (ALMA) \citep{alm15,aki16} has triggered hot debates on their origins. The suggested origins can be classified into the three categories; 1) (sub-)Jovian mass planets \citep{dip15,kan15,tam15,don15,jin16}, 2) change of dust properties \citep{zha15,oku16}, and 3) secular gravitational instabilities \citep{tak14}. 
Later, \citet{yen16} have found signs of gaps of molecular gas in the HCO$^+$ emission at radii of $\sim$28 and $\sim$69 au, coincident with the locations of the dust gaps. 
The coincident gas and dust gaps would favor the planetary origin of these gaps in the circumstellar disk. Subsequent ALMA observations in C$^{18}$O and $^{13}$CO at a 1$\arcsec$ resolution by \citet{yen17b} suggest that the mass of HL Tau is 1.8 $M_{\odot}$ from the observed Keplerian rotation, and that the outer envelope has an infalling velocity higher than the free-fall velocity, which could be due to the compression by the expanding shell driven by XZ Tau.
These studies show that HL Tau is a protostellar source associated with both the candidate disk of planet formation and the infalling protostellar envelope.

Recent observational studies of Class \RN{2} disks have found that the ratio of the disk gas mass derived from CO isotopologue lines to the disk dust mass is significantly lower than the canonical value of 100 \citep{ans16,lon17,mio17}. Estimations of disk gas masses using HD (1--0) have suggested the low CO abundances in a few protoplanetary disks \citep{mcc16,sch16}. The decrease of the CO abundance could be explained by the chemistry as follows. He$^+$ reacts with CO  to form C$^{+}$, and subsequently C$^{+}$ could react with other molecules and form less volatile ices, e.g., CO$_{2}$ and hydrocarbons \citep{aik97,ber14}. The disk around HL Tau is a candidate site of planet formation. Thus, HL Tau is an excellent target to study the physical and chemical conditions of sites of planet formation at the early phase of star formation in comparison with other Class \RN{1} and Class \RN{2} disks.

In this paper, we report our combined efforts of SMA observations around HL Tau and further analyses of the ALMA archival data. 
The large fields of view (FOV) of $\sim$1$\arcmin$ (8400 au) of our SMA observations at 1.3 mm and the ALMA HCO$^+$ observations at 2.9 mm enable us to derive the physical conditions of the extended protostellar envelope.
In addition, these observations with the shortest baseline lengths of $<$ 7.5 k$\lambda$ can detect structures on a scale larger than 12$\arcsec$ ($\sim$1700 au), twice larger than that in the ALMA $^{13}$CO and C$^{18}$O observation by \citet{yen17b}. 
We have also obtained HCO$^+$, $^{13}$CO and C$^{18}$O spectra of HL Tau with the IRAM 30 m telescope to measure the total flux. 
In this work, we study the physical and chemical conditions and the gas motions in the disk and envelope in HL Tau, and discuss our results in the context of evolution of gas and dust components in the disk and envelope. In Section 2 of the present paper, we describe our data reduction of the ALMA archival data and the details of our new observations of HL Tau with the SMA and IRAM 30 m telescope. In Section 3, we present the SMA results of the 1.3 mm dust-continuum, $^{13}$CO (2--1), C$^{18}$O (2--1), and SO (5$_6$--4$_5$) emission and the ALMA results of the HCO$^{+}$ (1--0) emission. The gas motions and physical conditions in the disk and envelope are discussed with comparisons of these ALMA and SMA results and our disk+envelope models in Section \ref{sec:ana}. Then we discuss implications from these results in Section \ref{discuss}. Section \ref{sum} provides a concise summary of our main results and discussion.

\section{Observations}
\subsection{HCO$^+$ Observations with ALMA}
The HCO$^+$ (1--0) emission in HL Tau was observed as a science verification of the ALMA Long Baseline Campaign (project code: 2011.0.00015.SV). The details of the observations were described by \citet{alm15}. The projected baseline lengths of the ALMA long baseline data range from 13.2 m to 15.1 km (3.93 k$\lambda$ $\sim$ 4533 k$\lambda$ at the HCO$^+$ frequency). We retrieved the calibrated data from the public archive, and followed the procedure of the imaging described in the archive\footnote{\url{https://casaguides.nrao.edu/index.php?title=ALMA2014_LBC_SVDATA}.} using Common Astronomy
Software Applications (CASA; \citet{mcm07}) of version 4.2.2. The image cube was made with a uv-taper of 200 k$\lambda$ (0$\farcs$11) and briggs weighting with a robust parameter of 0.5. These parameters were chosen  to optimize the angular resolution and sensitivity to meet the requirements of our scientific interest. The resultant beam size and rms noise level per channel are $\sim$1.0" and 2.4 mJy (0.35 K), respectively. The basic parameters of the map are summarized in Table \ref{table:imaging}.

\begin{deluxetable}{cccccccc}
\tabletypesize{\scriptsize}
\tablewidth{0pt}
\tablecaption{Parameters of the continuum and molecular-line velocity channel maps of the ALMA and SMA Observations.\label{table:imaging}}
\tablehead{
\colhead{molecular line}&\colhead{telescope}&\colhead{primary} &\colhead{rest}&\colhead{velocity} &\colhead{robust} &\colhead{beam size(\arcsec)} & \colhead{noise}\\
\colhead{/continuum}&\colhead{}&\colhead{beam(\arcsec)}&\colhead{frequency}&\colhead{resolution}&\colhead{position angle}&\colhead{(mJy Beam$^{-1}$)}\\
\colhead{}&\colhead{}&\colhead{}&\colhead{(GHz)}&\colhead{(km s$^{-1}$)}&\colhead{}&\colhead{}&\colhead{}
}
\startdata
HCO${^+}$ (1--0)&ALMA &   71\arcsec    &89.188518 &0.42&          0.5 &\makecell{1$\farcs$05$\times$0$\farcs$95\\(P.A.= -84$\degree$)} &2.4\\
$^{13}$CO (2--1)&SMA  & 55\arcsec      &220.398684&0.28 &          -0.5 &\makecell{1$\farcs$63$\times$1$\farcs$49\\(P.A.= -65$\degree$)} &60\\
C$^{18}$O (2--1) &SMA&55\arcsec  &219.560358 &0.28&          0.5 &\makecell{1$\farcs$78$\times$1$\farcs$59\\(P.A.= -76$\degree$)} &34\\ 
SO (5$_6$--4$_5$)&SMA&55\arcsec  &219.949433&0.28 &          0.5&\makecell{1$\farcs$95$\times$1$\farcs$75\\(P.A.= -73$\degree$)} &33\\
continuum&SMA&55\arcsec&225&...&0.5&\makecell{2$\farcs$23$\times$1$\farcs$95\\(P.A.= -78$\degree$)}&2.9\\
\enddata
\end{deluxetable}

\subsection{SMA Observations at 225 GHz}
The SMA observations at 225 GHz were conducted during the period from February to December of 2015 with the three different array configurations, the sub-compact, compact, and the extended configurations. The projected baseline lengths of the observations range from 9.6 m to 226 m (7.4 k$\lambda$ $\sim$ 174 k$\lambda$). The field center of the SMA observations is $04\hours31\minutes38\seconds.42$, $+18\degree13\arcmin57\farcs37$ (J2000). It is 0$\farcs$3 northwest of the field center of the ALMA observations, which is $04\hours31\minutes38\seconds.43$, $+18\degree13\arcmin57\farcs05$. This difference is negligible compared to the fields of view of these ALMA and SMA observations. 
We observed the CO (2--1), $^{13}$CO (2--1), C$^{18}$O (2--1), and SO (5$_6$--4$_5$) lines. 
In this paper, the results of $^{13}$CO, C$^{18}$O, and SO are presented.
The bandwidth of the observations in February was 4 GHz (USB) + 4 GHz (LSB)  using the ASIC correlator. It was expanded to 7GHz + 7GHz in the observing runs in October and December with the combination of the ASIC and SWARM correlators. 
However, because of the operational instability of the newly equipped SWARM correlator, the data obtained with SWARM were not used for imaging. 
The ASIC correlator consists of 48 chunks, and each chunk has a bandwidth of 104 MHz. 
 512 channels per chunk were assigned to the molecular lines, which provide a velocity resolution of 0.28 km s$^{-1}$ at the C$^{18}$O (2--1) frequency. 
The C$^{18}$O data observed on December 19 were discarded due to the disability of the chunk assigned to the C$^{18}$O line during the observation.
The visibility data were calibrated using the MIR IDL package (https://www.cfa.harvard.edu/rtdc/SMAdata/process/mir/) and imaged using the MIRIAD package \citep{sau95}. 
The C$^{18}$O and SO images were generated with briggs weighting with a robust parameter of 0.5, while the $^{13}$CO image with a robust parameter of -0.5 to suppress the sidelobes. 
The 225 GHz continuum data were obtained by averaging the line-free chunks of the ASIC correlator, and the data from the upper and lower sidebands were combined to improve the signal-to-noise ratio. 
The continuum image was generated with a robust parameter of 0.5.
The details of the observations are summarized in Table \ref{table:observation}, and the synthesized beams and noise levels of the images in Table \ref{table:imaging}.

\begin{deluxetable}{lcccc}
\tabletypesize{\footnotesize}
\tablecolumns{5}
\tablewidth{0pt}
\tablecaption{Parameters of the SMA Observations in 2015.\label{table:observation}}
\tablehead{parameter&\multicolumn{4}{c}{value}}
\startdata
date&Feb 1&Oct 29&Dec 19&Dec 27\\
weather (opacity $\tau_{\rm 225GHz}$)&0.02--0.12&0.07--0.08&0.1--0.25&0.07--0.15\\
Tsys (K)&150--250&150--220&200--400&150--400\\
configuration&extended&subcompact&compact&compact\\
antenna number&6&6&8&8\\
bandpass calibrator&3c279&3c273&\multicolumn{2}{c}{3c279}\\
flux calibrator&Uranus&\multicolumn{3}{c}{Callisto}\\
gain calibrator&3c120 \& 0423-013&\multicolumn{3}{c}{3c120 \& 0510+180}\\
flux of 3c120 (Jy)&3.3&3.3&3.4&3.8\\
flux of 0423-013 (Jy)&1.1&&&\\
flux of 0510+180 (Jy)&&2.7&3.4&2.8\\
\enddata
\end{deluxetable}

 \subsection{IRAM 30 m observations}
Mapping observations of HL Tau with the IRAM 30 m telescope were conducted with the EMIR receiver and the FTS spectrometer on March 28 and 29 in 2017. The total on-source time is 4.7 hours. The  $^{13}$CO (2--1), C$^{18}$O (2--1), SO (5$_6$--4$_5$), and HCO$^{+}$ (1--0) lines were observed simultaneously at a spectral resolution of 50 kHz. The data were reduced with CLASS (http://www.iram.fr/IRAMFR/GILDAS). The noise level is 0.2--0.3 K in main beam temperature at a velocity resolution of 0.07 km s$^{-1}$. The main beam efficiency is 0.61 at 220 GHz and is 0.75 at 89 GHz. 
In this paper, we present the spectra of the $^{13}$CO (2--1), C$^{18}$O (2--1), SO (5$_6$--4$_5$) and HCO$^{+}$ (1--0) emission at the position of HL Tau.
The details of the observations and the results will be presented in our forthcoming paper (Yen et al. in prep.).
\section{Results}

\subsection{1.3 mm Continuum Emission}\label{cont_sec}
\begin{figure}
\epsscale{1} 
\plotone{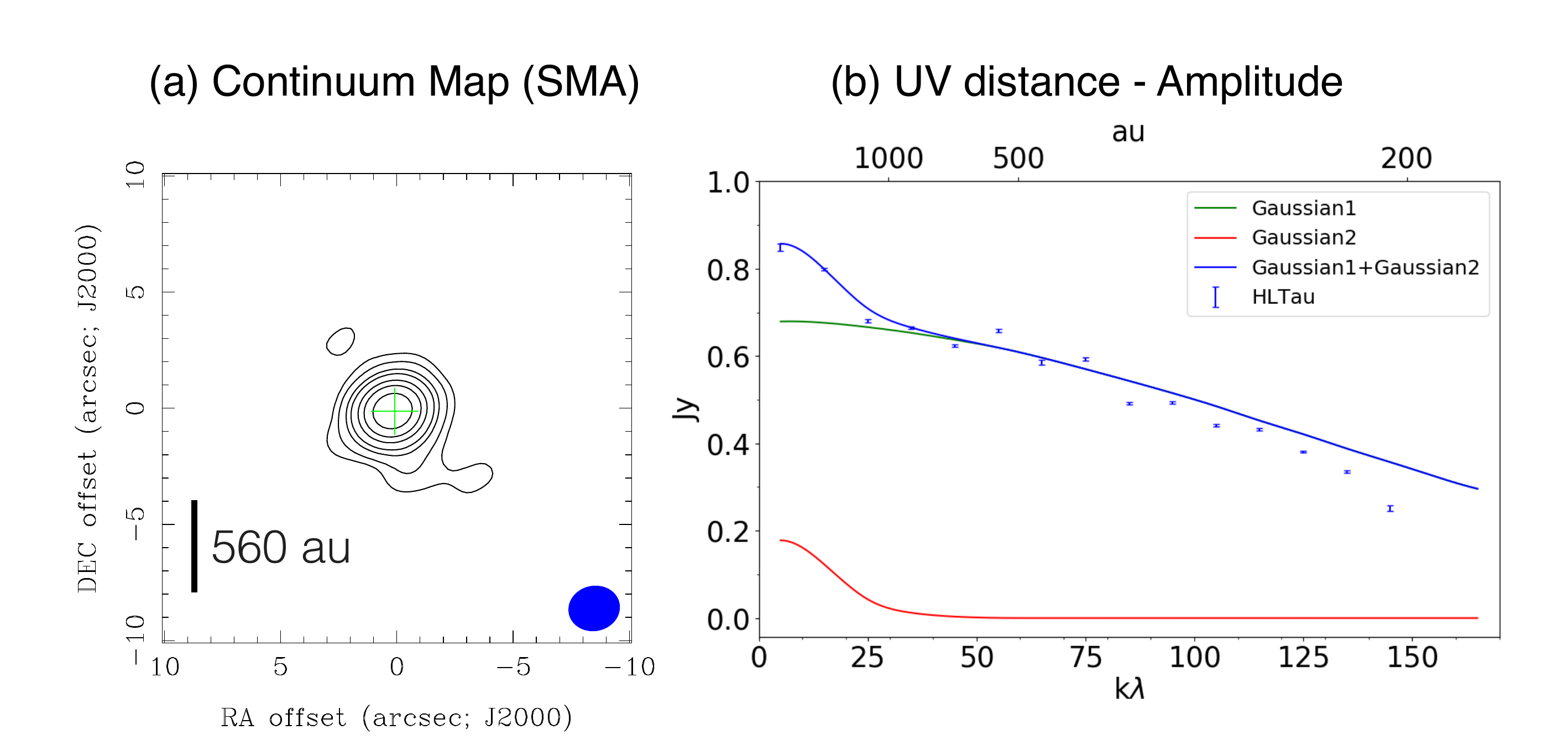}
\caption{(a) 1.3 mm continuum image observed with the SMA. 
The coordinates are the offsets from the peak position of the compact component determined from the two-component Gaussian fitting to the visibility data, which is regarded as the protostellar position.
 Contour levels are 7$\sigma$, 14$\sigma$, 28$\sigma$, 42$\sigma$, 70$\sigma$, 100$\sigma$, 150$\sigma$, where 1 $\sigma$ is 2.9 mJy beam$^{-1}$. The beam size is 2$\farcs$23 $\times$ 1$\farcs$95, as shown at the bottom right corner. A green cross indicates the protostellar position. (b) Visibility amplitude versus {\it uv} distance plot of the continuum emission with 1$\sigma$ error bars. Green  and red curves show profiles of the two fitted Gaussian components to the observed visibilities. The blue curve is the sum of the two components. 
}
\label{Fig:cont}
\end{figure}

Figure \ref{Fig:cont}a shows the 1.3 mm continuum map obtained with the SMA. 
The spatial structure is more clearly seen in the visibility amplitude profile shown in Figure \ref{Fig:cont}b, which exhibits two components with different spatial scales.
In addition to a central compact component that dominates the flux at long {\it uv}-distance of $>$ 50 k$\lambda$, a spatially extended component is observed as an excess emission at short {\it uv}-distance of $<$ 20 k$\lambda$.
 The visibility data were fitted with two elliptical Gaussian components using the MIRIAD task {\it uvfit}. 
 The compact component has a size of 0$\farcs$84$\times$0$\farcs$45 (118$\times$63 au), and an extended component has a size of 6$\farcs$5$\times$4$\farcs$3 (910$\times$600 au), on the assumption of a common peak position of the two components.
Detailed parameters of these two components are given in Table \ref{table:gaussians}.  
The flux of the compact component is $\sim$ 78$\%$ of the total 1.3 mm flux. 
The position angle of the compact component is 138$\degree$. 
The inclination angle derived from the ratio of the major and minor axes is 58$\degree$, assuming that the compact component is a geometrically thin disk.
The position and inclination angels of the compact component are consistent with those of the circumstellar disk observed with ALMA \citep{alm15} and CARMA \citep{kwo11} at higher angular resolutions in 1.3 mm.  
The flux of the compact component, 680 mJy, is also consistent with the previous measurements of 700 mJy \citep{kwo11}, if we take into account the absolute flux uncertainty of $\sim$10 \%. The flux of the extended component is measured to be $\sim$190 mJy. 
Because the total flux (i.e. compact + extended components) of $\sim$870 mJy is consistent with the flux of $\sim$880 mJy measured with the single-dish telescope at an angular resolution of 11$\arcsec$ by \citet{bec90}, the missing flux of the extended component is likely negligible.  
Since the dust emission from the disk is partially optically thick at 1.3 mm \citep{jin16,pin16}, to estimate the disk dust mass requires detailed radiative transfer modeling.
The previous work using the 7 mm data estimated the disk dust mass to be 1--3$\times 10^{-3}$ $M_{\odot}$ \citep{car16}.
Another estimation with the 1.3 mm and 2.7 mm data also provided the consistent number of 1.35$\times 10^{-3}$ $M_{\odot}$ \citep{kwo11}.
Our measured 1.3 mm flux agrees well with \citet{kwo11}.
Given the above estimates, in the present paper we adopted the disk dust mass of 1--3$\times 10^{-3}\ M_{\odot}$. 

We estimate the dust mass of the extended component with the following equation,
\begin{equation}
M_{\rm dust} =\frac{F_\nu d^2}{\kappa_\nu B(T_{\rm dust})},
\end{equation}
where $F_{\nu}$ is the observed flux density at the frequency $\nu$, $d$ is the distance to the source, $\kappa_\nu$ is the dust mass opacity, and $B(T_{\rm dust})$ is the Planck function at the dust temperature $T_{\rm dust}$.
In the envelope, $\kappa_\nu$ is assumed to be 0.899 cm$^2$ g$^{-1}$ \citep{oss94} and $T_{\rm dust}$ is assumed to be 15 K, 
which is the kinetic temperature estimated from the C$^{18}$O (3--2) and (2--1) intensity ratio (Yen et al. in prep.). 
This $\kappa_\nu$ value is the dust-only mass opacity, and the mass derived here is the dust mass rather than dust+gas mass. 
The estimated dust mass in the envelope is 1.2$\times10^{-3}$ $M_{\odot}$. 

With the estimated envelope and disk masses from the continuum emission, the envelope-to-disk mass ratio in HL Tau is derived to be 0.4--1.2, higher than the ratio of 0.1 estimated by \citet{she18}. This difference could be due to the different $uv$-sampling. The envelope mass could be underestimated in \citet{she18} because they only adopted the data with the baseline length of $>$16 k$\lambda$.

\begin{deluxetable}{lccc}
\tabletypesize{\footnotesize}
\tablecolumns{3}
\tablewidth{0pt}
\tablecaption{Parameters of the two Gaussian components obtained from the fitting to the visibility data of the 1.3 mm continuum emission \label{table:gaussians}}
\tablehead{&compact Gaussian&extended Gaussian}
\startdata
flux density (Jy)&0.68&0.19\\
offset in RA, DEC (arcsec)&(0.18,-0.11)&(0.18,-0.11)\\
major, minor axes (arcsec)&(0.84,0.45)&(6.5,4.3)\\
position angle (degrees)&-42&31\\
\enddata
\end{deluxetable}

\subsection{Molecular line maps}
\begin{figure}
\epsscale{1} 
\plotone{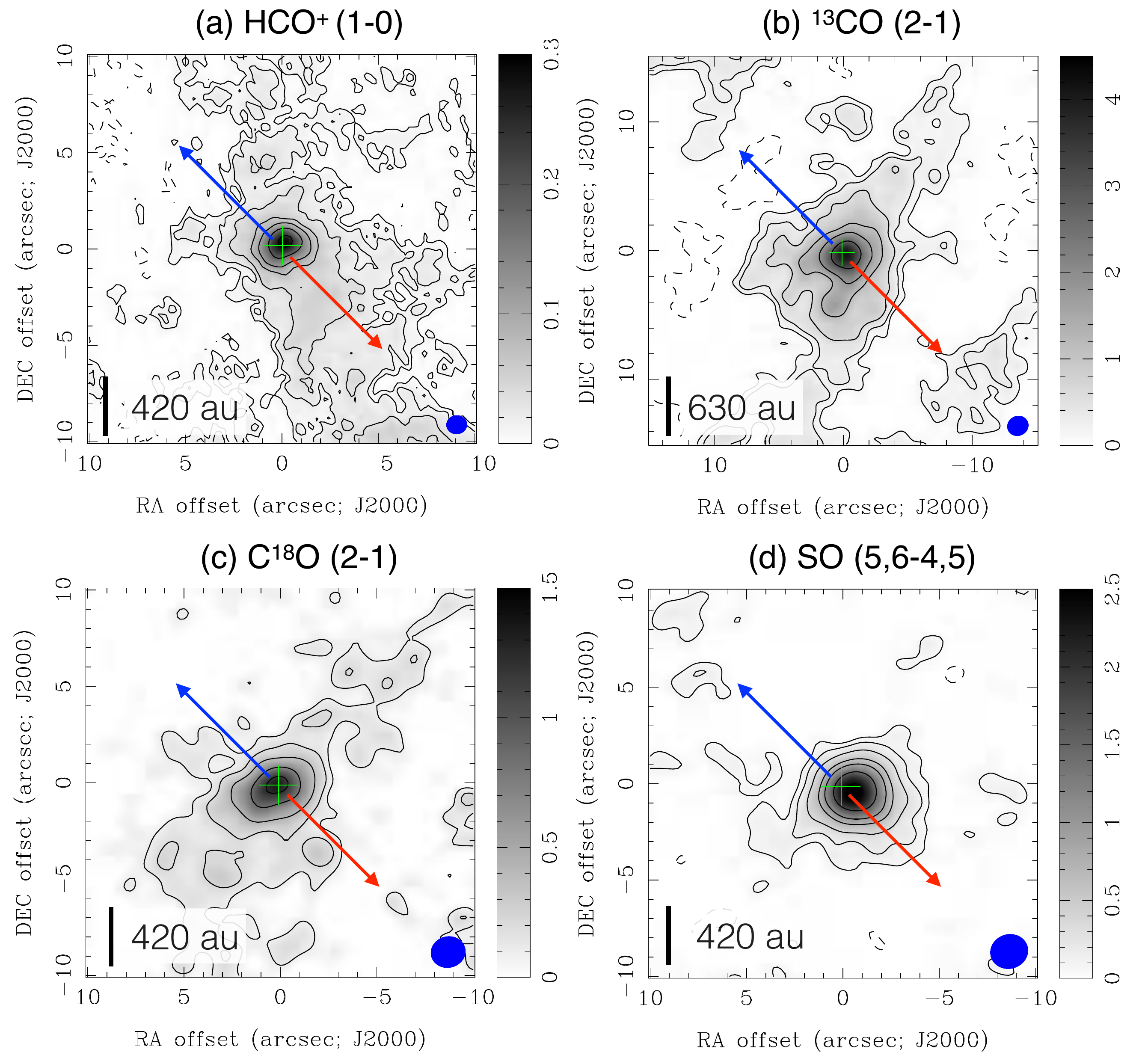}
\caption{Moment 0 maps of the (a) ALMA HCO$^+$(1--0), (b) SMA $^{13}$CO(2--1), (c) C$^{18}$O(2--1), and (d) SO(5$_6$--4$_5$) lines. The integrated velocity ranges of the HCO$^+$, $^{13}$CO, C$^{18}$O, and SO lines are $V_{\rm LSR}$ of 1.25--13 km s$^{-1}$, 3.5--11.75 km s$^{-1}$, 3.75--12.25 km s$^{-1}$, and 5.5--11 km s$^{-1}$, respectively. Blue ellipses at the bottom right corners denote the synthesized beams (Table \ref{table:imaging}). Green crosses denote the protostellar position of HL Tau. Contour levels of the maps are -4$\sigma$, 2$\sigma$, 4$\sigma$, 8$\sigma$, 16$\sigma$, 24$\sigma$, 40$\sigma$, 64$\sigma$, where 1$\sigma$ of the HCO$^+$, $^{13}$CO,  C$^{18}$O, SO maps are 4.2, 73, 50, and 40 mJy beam$^{-1}$ km s$^{-1}$, respectively. 
}
\label{Fig:moment0}
\end{figure}

Figure \ref{Fig:moment0} shows the moment 0 maps of HL Tau in the four molecular lines, the HCO$^+$ (1--0) line observed with the ALMA, and the $^{13}$CO (2--1), C$^{18}$O (2--1), and SO (5$_6$--4$_5$) lines observed with the SMA. 
Since the spatial extent of the $^{13}$CO emission is larger than that of the other lines, the area shown in Figure \ref{Fig:moment0}b is larger than that of the other panels.
These maps show that the molecular-gas structures consist of the central compact components surrounded by the extended emission.

The extended component of the $^{13}$CO (2--1) emission is elongated from southeast to northwest which is consistent with the orientation of the $^{13}$CO (1--0) emission observed by \citet{hay93}.
Since this elongation is parallel to the major axis of the continuum disk \citep{alm15,kwo11}, and is perpendicular to the outflow and jet \citep{mun90,lum14}, most of the extended $^{13}$CO emission observed with SMA likely traces the flattened protostellar envelope.
The presence of the flattened structure is also supported by the C$^{18}$O map, which shows a clear elongation in both the compact and extended components.
Assuming that the C$^{18}$O emission traces an inclined disk-like structure, the position and inclination angles are derived to be 119$\degree$ and 49$\degree$, respectively.
These numbers agree with the position and inclination angles of the continuum disk, 138$\degree$ and 47$\degree$, observed at the higher angular resolution \citep{alm15}.
This suggests that the C$^{18}$O emission observed with SMA also traces the flattened envelope and the disk, and the envelope and the disk are likely coplanar.
On the other hand, the extended component of the HCO$^+$ (1--0) emission exhibits elongation along the axis of the associated jets and outflows, suggesting the contamination from the outflow.

In contrast to the $^{13}$CO, C$^{18}$O and HCO$^+$ images having a compact component centered at the continuum peak, the center of the compact component in the SO map is shifted to the southwest (Figure \ref{Fig:moment0}d).
The center of the SO emission derived from the two-dimensional Gaussian fitting is 0$\farcs$67 south and 0$\farcs$43 west from the protostellar position,
which corresponds to a positional offset of $\sim$110 au from the protostar. As the direction toward the SO emission peak matches the direction of the redshifted molecular outflow, the SO emission likely has the contamination from the outflow.

 \subsection{Velocity Structure of HL Tau}
\begin{figure}
\epsscale{0.9}
\plotone{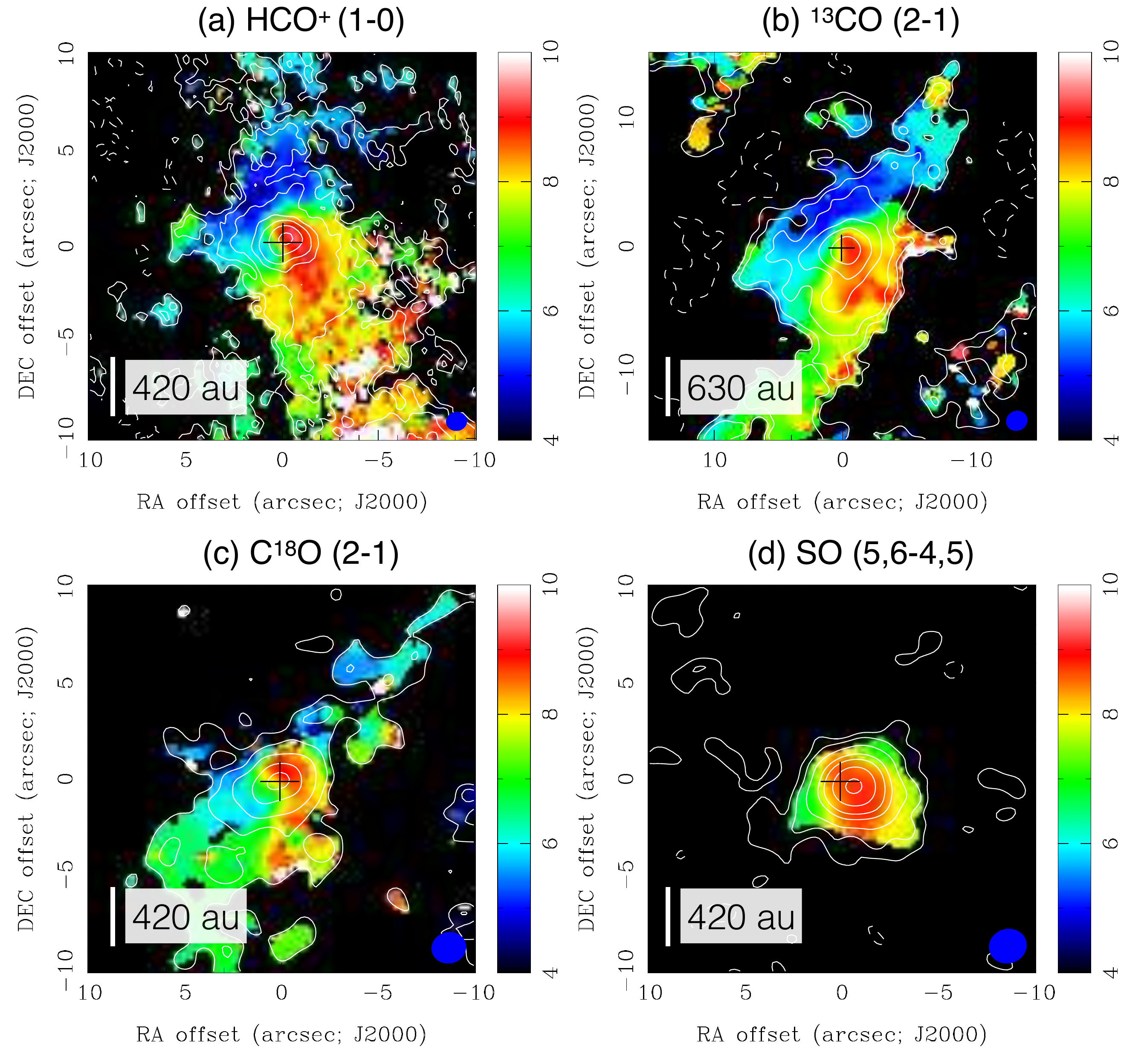}
\caption{Moment 1 maps (color) overlaid with the moment 0 maps (white contours) of the (a) HCO$^+$, (b) $^{13}$CO, (c) C$^{18}$O, and (d) SO lines. Blue ellipses at the bottom right corners and black crosses denote the relevant synthesized beams and the protostellar position of HL Tau, respectively. Contour levels are the same as those in Figure \ref{Fig:moment0}.
}
\label{Fig:moment0+1}
\end{figure}

Figure \ref{Fig:moment0+1} shows the moment 1 maps (color) overlaid with the moment 0 maps (white contours) of the HCO$^+$, $^{13}$CO, C$^{18}$O, and SO emission.
The moment 1 maps of the HCO$^+$, $^{13}$CO, and C$^{18}$O reveal that the central compact and surrounding extended components have different velocity gradients.
In the extended component of the $^{13}$CO emission, there is a velocity gradient from northeast (blueshifted) to southwest (redshifted), which is along the minor axis of the continuum disk.
The emission extending to the northwest is blueshifted, while that extending to the southeast has a velocity close to the systemic velocity of $V_{\rm LSR}$ = 7.0 km s$^{-1}$ \citep{alm15}.
In the central compact component, the $^{13}$CO emission shows a velocity gradient along the east--west direction with the blueshifted part to the east and the redshifted part to the west.
The overall velocity structures seen in the extended component of the C$^{18}$O emission is consistent with that in the extended $^{13}$CO component.
The extended C$^{18}$O component in the northwest is blueshifted and that of the southeast is close to the systemic velocity.
In addition, the redshifted C$^{18}$O emission is extended to the south of the compact component.
These outer extended components were not clearly seen in the C$^{18}$O image obtained with the ALMA \citep{yen17b}.
On the other hand, the velocity gradient of the central compact component of the C$^{18}$O emission is different from that seen in the $^{13}$CO emission.
It is from southeast (blueshifted) to northwest (redshifted), which is along the major axis of the disk.
This difference is probably because the velocity gradient of the $^{13}$CO emission is more contaminated by the redshifted component extending to the south.
The HCO$^+$ emission exhibits similar velocity gradients to those of the $^{13}$CO and C$^{18}$O emission.
The outer extended part shows a velocity gradient  along the northeast--southwest direction, with the blueshifted part to the northeast and the redshifted part to the southwest. The central compact component has a different velocity gradient along the southeast--northwest direction, which is along the disk major axis, with the blueshifted part to the southeast and the redshifted part to the northwest, the same as that in the C$^{18}$O emission.
Different from the other three lines, the SO emission does not show a clear velocity gradient. 
Most of the SO emission is redshifted. %, although parts of the SO emission around the systemic velocity is seen in the velocity channel maps. (SO channel maps are not shown.)

\begin{figure}
\epsscale{1} 
\plotone{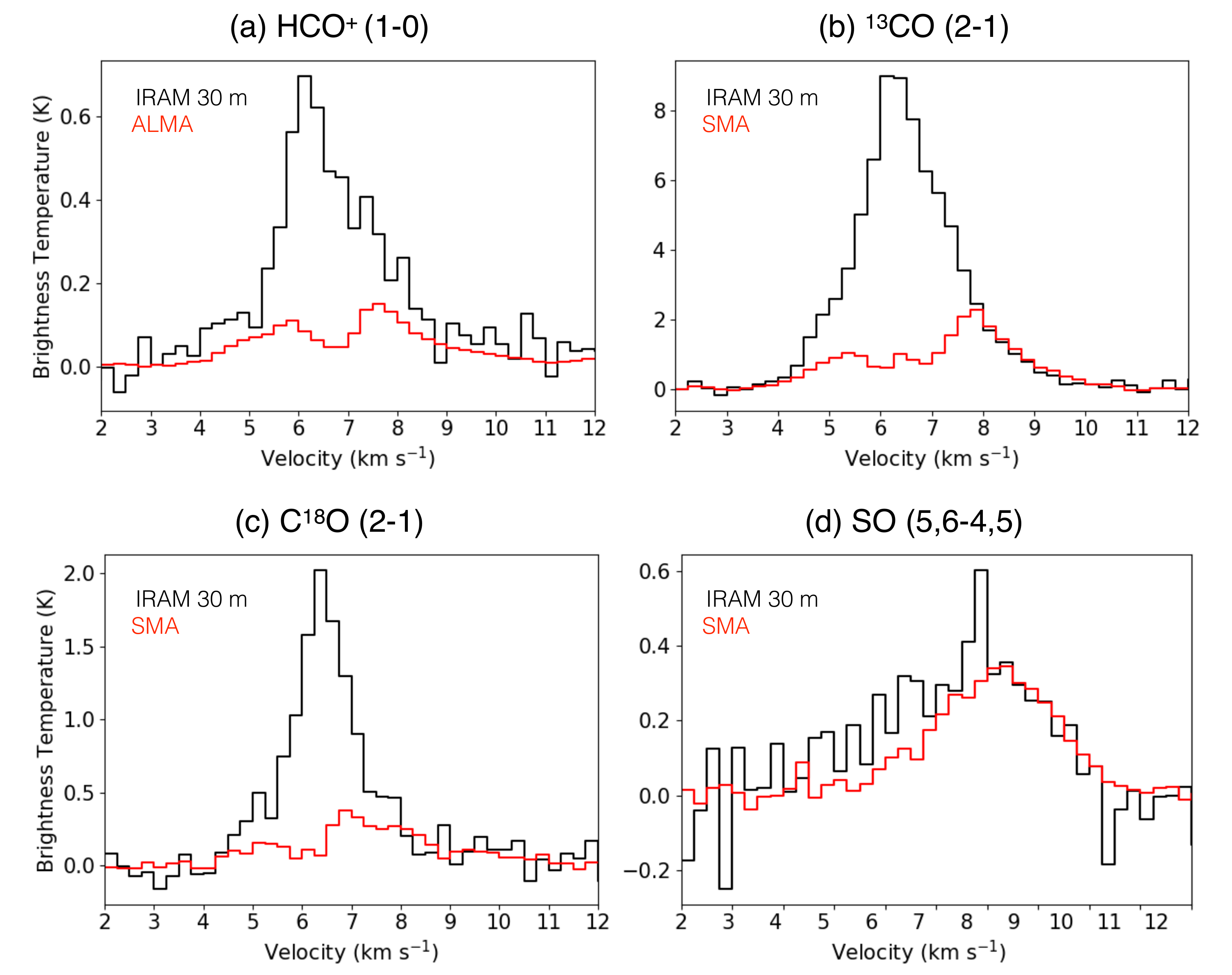}
\caption{Comparisons of the spectra obtained with the IRAM 30 m telescope (black) and with the interferometers (red), (a)  HCO$^+$ (1--0) with the ALMA, (b--d) $^{13}$CO (2--1), C$^{18}$O (2--1), and SO (5$_6$--4$_5$) with the SMA at the protostellar position of HL Tau. }
\label{Fig:spectral}
\end{figure}

Since the images obtained with interferometers often suffer from the effect of  missing flux, we have estimated the missing flux of each molecular line using the spectra obtained with the IRAM 30 m telescope.
The SMA and ALMA maps were primary beam corrected and convolved with the same beam size of the IRAM 30 m observations. The spectra extracted from the convolved images were compared with the IRAM 30 m spectra.
As shown in Figure \ref{Fig:spectral}, the flux tends to be well recovered at the redshifted velocity of $V_{\rm LSR} >$ 8 km s$^{-1}$.
Especially, the $^{13}$CO flux is well recovered in this velocity range, implying that the redshifted $^{13}$CO emission is almost free from the missing flux.
On the other hand, the missing flux is significant at the blueshifted velocity.
At the velocity of 4--8 km s$^{-1}$, the missing flux is $\sim$80$\%$ in the $^{13}$CO and C$^{18}$O emission observed with SMA and is more than 80$\%$ in the HCO$^{+}$ emission observed with ALMA, likely caused by the presence of the large-scale filamentary structure at these velocities \citep{wel00}.
In contrast, the SO emission has no significant missing flux over the entire velocity range, suggesting that the distribution of the SO emission in HL Tau is compact as observed with SMA.

\begin{figure}
\epsscale{0.7} 
\plotone{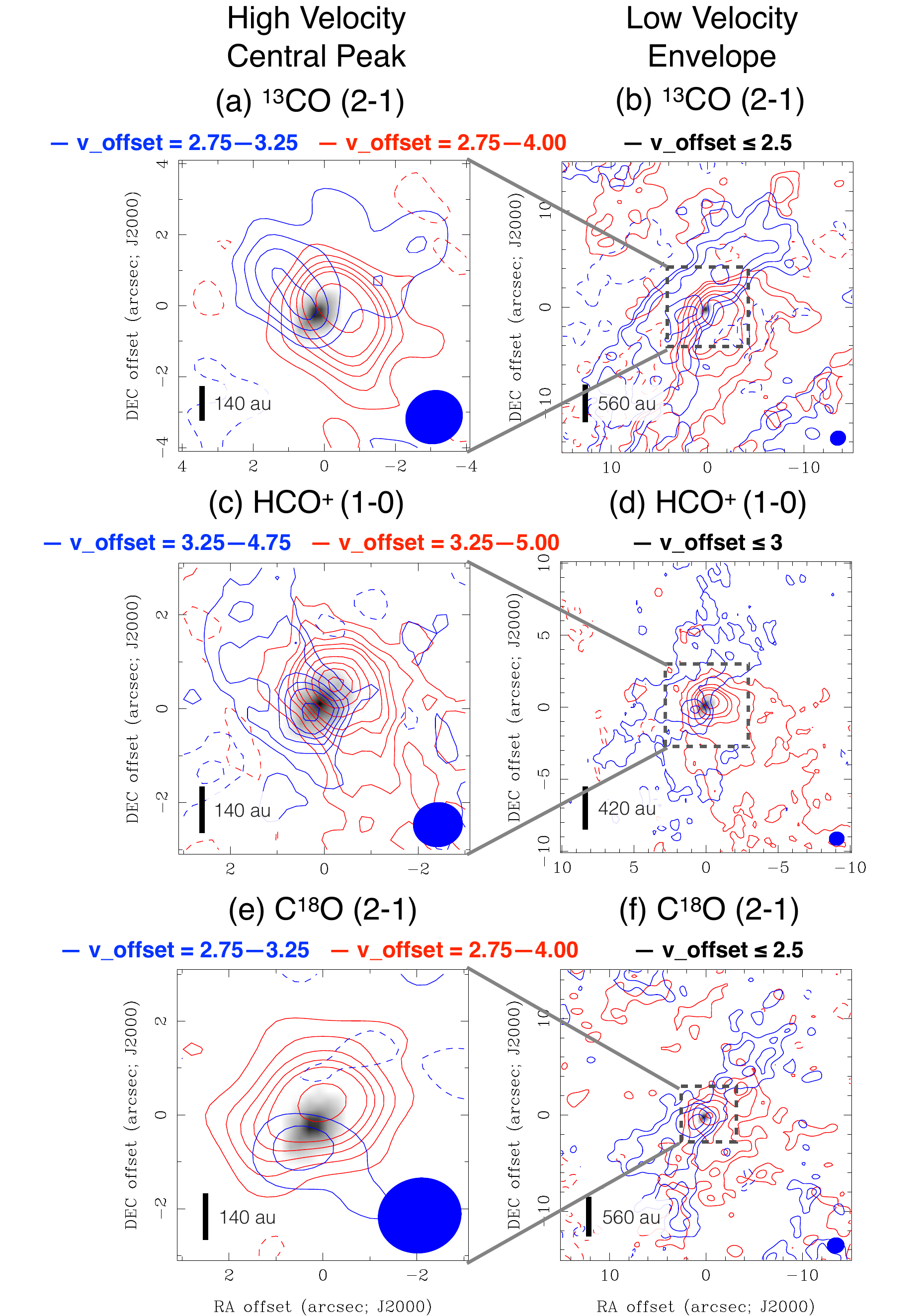}
\caption{Maps of the $^{13}$CO, HCO$^+$, and the C$^{18}$O emission integrated over two different velocity ranges superposed on the 2.9 mm continuum image (grey scale) obtained with the ALMA \citep{alm15}. 
The integrated velocity ranges with respect to the systemic velocity of $V_{\rm LSR}$ = 7.0 km s$^{-1}$ are listed above each panel in units of km s$^{-1}$. 
The integrated velocity ranges of the high-velocity blue- and redshifted emission are shown as blue and red numbers, respectively, 
and those of the low-velocity emission as black numbers.
Blue and red contours present the blue- and redshifted emission, respectively.
 The contour levels of the maps (a), (c), (e) are -8$\sigma$, -4$\sigma$, 2$\sigma$, 4$\sigma$, 8$\sigma$, 12$\sigma$, 16$\sigma$, 24$\sigma$, 32$\sigma$, 40$\sigma$, 48$\sigma$. The contour levels of the maps (b), (f) are -4$\sigma$, 2$\sigma$, 4$\sigma$, 8$\sigma$, 16$\sigma$, 24$\sigma$, 40$\sigma$. The contour levels of the map (d) are -5$\sigma$, 5$\sigma$, 10$\sigma$, 20$\sigma$, 30$\sigma$, 40$\sigma$, 70$\sigma$. 1$\sigma$ levels of blueshifted components of (a), (c), (e) are 26, 1.8, and 15 mJy beam$^{-1}$ km s$^{-1}$, respectively. 1$\sigma$ levels of redshifted components of  (a), (c), (e) are 37, 1.8, and 21 mJy beam$^{-1}$ km s$^{-1}$, respectively. 1$\sigma$ levels of both blue- and redshifted components of (b), (d), (f) are 50, 2.2, and 28 mJy beam$^{-1}$ km s$^{-1}$, respectively.
 The blue ellipses at the bottom right corners present the beam sizes (Table \ref{table:imaging}). 
 }
\label{Fig:difvel}
\end{figure}

Figure \ref{Fig:difvel} presents the $^{13}$CO, HCO$^+$, and C$^{18}$O moment 0 maps of two different velocity ranges.
Figures \ref{Fig:difvel}a, 5c, and 5e show the blue- and redshifted components at the high velocity. 
The high-velocity ranges are the velocity offsets ($\Delta V$) of $|\Delta V| > 2.5$ km s$^{-1}$ for the $^{13}$CO and C$^{18}$O lines and $|\Delta V| > 3.0$ km s$^{-1}$ for the HCO$^+$ line with respect to the systemic velocity of $V_{\rm LSR}$ = 7.0 km s$^{-1}$. 
Figures \ref{Fig:difvel}b, 5d, and 5f present the low-velocity components with $|\Delta V| \leq 2.5$ km s$^{-1}$ for the $^{13}$CO and C$^{18}$O lines and $|\Delta V| \leq 3.0$ km s$^{-1}$ for the HCO$^+$ line.
Figure \ref{Fig:difvel} shows that the high-velocity emission mainly traces the central compact component.
In the high-velocity HCO$^+$ and C$^{18}$O emission, the blueshifted and redshifted components are located to the southeast and northwest, respectively.
The peaks of the blue- and redshifted emission are located along the major axis of the dusty disk traced by the continuum emission, which is shown in grey scale.
On the other hand, the high-velocity $^{13}$CO emission shows a different pattern with the blueshifted part to the northeast and the redshifted part to the southwest. 
The same velocity pattern is also seen in the lower level contours of the high-velocity HCO$^+$ emission. 
This northeast-southwest velocity gradient is likely caused by the infalling envelope and the contamination of the outflow.
The peaks of the blue- and redshifted $^{13}$CO emission are located along the east-west direction as seen in the moment 1 map presented in Figure 3b.

The low-velocity emission is spatially extended, 
and the three lines show a similar velocity pattern at the low velocity.
The blueshifted emission is in the northeast, and the redshifted emission in the southwest of the continuum source.
In addition, the blueshifted emission is elongated along the northwest-southeast direction, as in the case of the blueshifted CO (1--0) emission \citep{alm15,kla16}.  
The orientations of the low-velocity blue- and redshifted components in the vicinity of the central source are consistent with those of the high-velocity components, 
along the east--west direction in the $^{13}$CO emission and along the southeast-northwest in the HCO$^+$ and C$^{18}$O emission.
 
\section{Analyses}\label{sec:ana}
\subsection{Gas Motions of the Circumstellar Material around HL Tau}\label{gas motion}
\subsubsection{Position--Velocity Diagram}
As described in the last subsection, two distinct,
almost orthogonal velocity gradients are identified around HL Tau.
One is the northwest (redshift) -- southeast (blueshift) gradient in the inner 100 au region at the high velocity of $|\Delta V| > 3.0$ km s$^{-1}$. 
The other is the northeast (blueshifted) -- southwest (redshifted) gradient in the outer region with a radius larger than 300 au at the low velocity of $|\Delta V| < 3.0$ km s$^{-1}$.
The direction of the velocity gradient in the inner region is consistent with the major axis of the disk observed in the dust continuum.
Thus, this velocity gradient most likely traces the Keplerian rotation in the disk. 
Such a velocity gradient originated from the Keplerian rotation has also been reported by \citet{alm15} and \citet{yen17b}.
On the other hand, the velocity gradient in the outer region likely traces the gas motion of the envelope surrounding the disk. 
Since the outflow associated with HL Tau is blueshifted to the northeast and redshifted to the southwest \citep{lum14}, the northeastern part of the disk and the flattened envelope is the far side, and the southwestern part is the near side. 
Therefore, the blueshifted emission on the far side and the redshifted emission on the near side likely trace the infalling gas motion in the flattened envelope.

\begin{figure}
\epsscale{0.8} 
\plotone{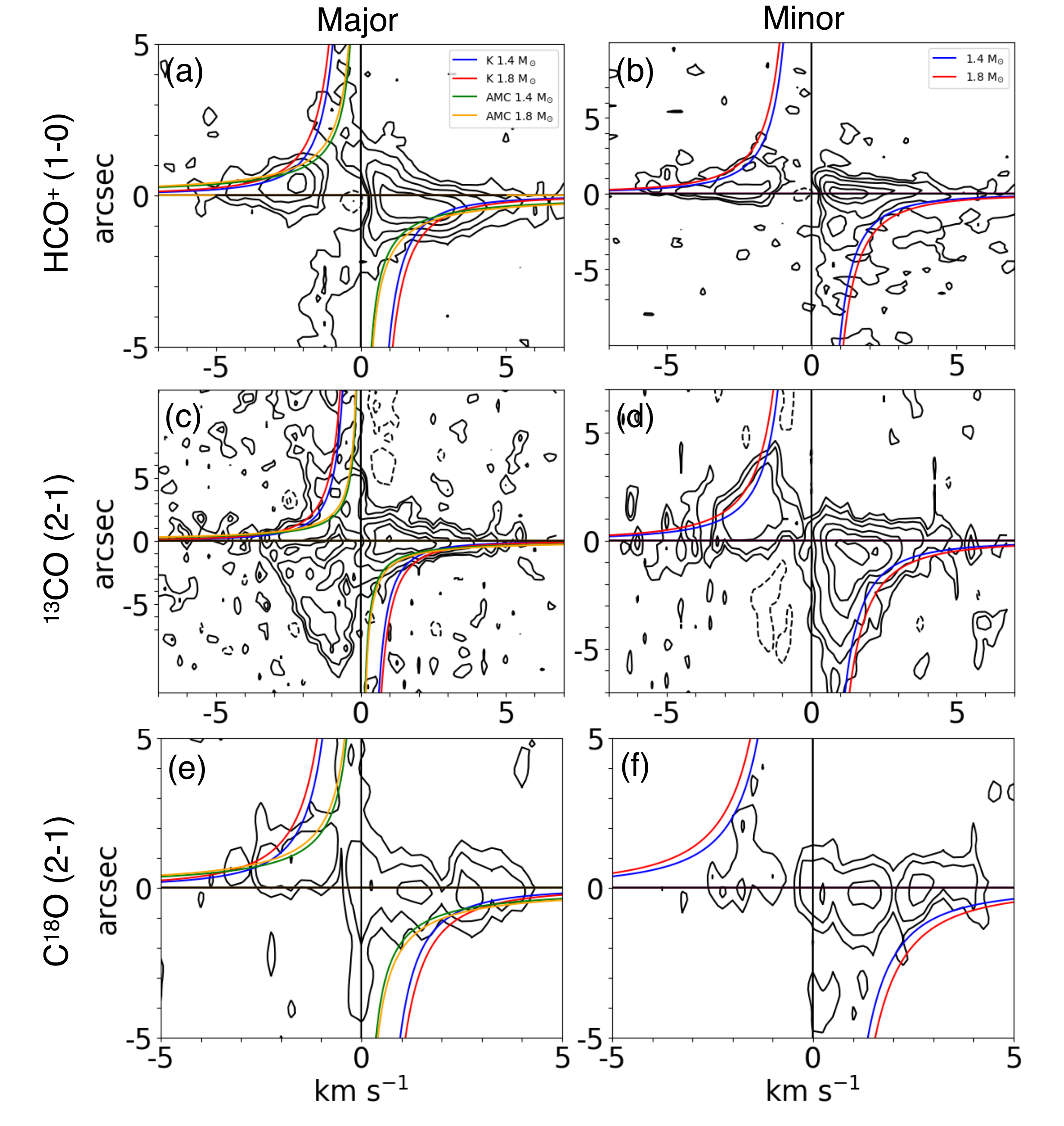}
\caption{Position -- Velocity (P--V) diagrams of the HCO$^+$ (1--0), $^{13}$CO (2--1), and the C$^{18}$O (2--1) emission along the major (left panels) and minor (right) axes. Contour levels are -4$\sigma$, 2$\sigma$, 4$\sigma$, 8$\sigma$, 16$\sigma$, 24$\sigma$, 40$\sigma$, where the 1$\sigma$ levels are listed in Table \ref{table:imaging}.
In the left panels, blue, red, green, and orange curves denote Keplerian rotation curves with the central stellar masses of 1.4 and 1.8 $M_{\odot}$ and rotation curves with the conserved specific angular momenta that connect to the Keplerian rotation curves with the central stellar masses of 1.4 and 1.8 $M_{\odot}$ at the radius of 100 au. In the right panels, blue and red curves denote free-fall curves toward the central stellar masses of 1.4 and 1.8 $M_{\odot}$. In Figure (a), the velocity profiles of the Keplerian rotation and angular momentum conservation with different central stellar masses are denoted as K and AMC, respectively.}
\label{Fig:PVmap}
\end{figure}

Figure \ref{Fig:PVmap} shows Position--Velocity (P--V) diagrams of the
HCO$^{+}$ (1--0), $^{13}$CO (2--1), and C$^{18}$O (2--1) emission along
the disk major and minor axes. In the P--V diagrams along the major axis,
Keplerian rotation curves with two different central stellar masses
($\equiv M_{\star} = 1.4$ and 1.8 $M_{\odot}$) are drawn, which are expressed as,
\begin{eqnarray}\label{keprot}
v_{\rm kep}(r)=\sqrt{\frac{GM_{\star}}{r}},
\end{eqnarray}
where $G$ is the gravitational constant and $r$ radius from the central star.
\citet{alm15} have reported the central protostellar mass of 1.3 $M_{\odot}$
from the observed disk rotation, while \citet{yen17b} have reported a
higher value of 1.8 $M_{\odot}$. 
By fitting our simple Keplerian disk model to the ALMA velocity channel maps of the HCO$^{+}$ (1--0) emission, the central
stellar mass is estimated to be 1.4 $M_{\odot}$ (see the next subsection). 
The comparison between the these Keplerian curves and the P--V diagrams along the major axis of the three molecular lines shows that the velocity structures at the high velocity of $|\Delta| > 3.0$ km s$^{-1}$ can be explained with the Keplerian rotation of the central stellar masses of 1.3--1.8 $M_{\odot}$.

At the high velocity, the P--V diagrams along the minor axis do not exhibit
any noticeable velocity gradients, consistent with the expectation from the Keplerian rotation.
As discussed above, at the lower velocity of $|\Delta V| < 3.0$ km s$^{-1}$ there exists an infalling
envelope surrounding the central Keplerian disk, and the radial infalling
motion in the flattened envelope is expected to show a velocity gradient along the
minor axis. If the infalling motion is free-fall, the velocity structure can be expressed as
\begin{eqnarray}
v_{\rm freefall}(r)=\sqrt{\frac{2GM_{\star}}{r}}.
\end{eqnarray}
The free-fall curves with the two different central stellar masses
($M_{\star} = 1.4$ and 1.8 $M_{\odot}$) are drawn in the P--V diagrams
along the minor axes. The P--V diagrams of the HCO$^{+}$ (1--0) and
$^{13}$CO (2--1) emission exhibit a
trend that the outer emission components are located at the lower
velocity, consistent with the free-fall curves.

In the infalling region outside the Keplerian disk, if the envelope rotation has a conserved specific angular momentum \citep{nak00}, 
its rotational profile can be described as,
\begin{eqnarray}
v_{\rm rot}(r)=\frac{R(100~{\rm au}) \times v_{\rm kep}(100~{\rm au})}{r}.
\end{eqnarray}
Here, the angular momentum is assumed to be the same as that of the
Keplerian rotation at the outermost radius of the disk (100 au). These
rotation curves are also drawn in the P--V diagrams along the major axis.
These curves better explain the velocity structure at radii larger than 1$\arcsec$--2$\arcsec$ in the P--V diagrams, compared to the Keplerian rotation curves.

From their ALMA Cycle 3 observations of HL Tau in the $^{13}$CO (2--1)
and C$^{18}$O (2--1) emission and the detailed modeling, \citet{yen17b}
have suggested that the gas kinematics in the protostellar envelope is rather complicated,
and that a simple kinematical model of the infalling and rotating flattened envelope
cannot fully explain the observed velocity structure. \citet{yen17b} have proposed
that the envelope kinematics is better explained when the influence of
the expanding shell driven by the neighboring young star, XZ Tau, on the envelope kinematics is considered.
The detailed envelope kinematics is beyond the scope of the present paper.
Instead, 
in the following we present our Keplerian disk model fitting to constrain the central stellar mass, 
and we perform a detailed modeling of the physical conditions of the HL Tau disk and envelope.

\subsubsection{Keplerian Disk Modeling}

To obtain quantitative parameters of the Keplerian disk for
the subsequent modeling of the physical conditions of
the disk and envelope, we performed
minimum $\chi^2$ fitting of the geometrically-thin Keplerian disk model
to the ALMA velocity channel maps of the HCO$^{+}$ (1--0) line at the
high-velocity of $|\Delta V| > 3$ km s$^{-1}$.
The HCO$^{+}$ (1--0) data are chosen for this fitting because
the critical density of the HCO$^{+}$
emission is an order of magnitude higher
than that of the $^{13}$CO (2--1) emission, suggesting less contamination
from the ambient materials, and
the signal-to-noise ratio of the C$^{18}$O (2--1) velocity
channel maps is not high enough to perform such a fitting.
Our fitting procedure is essentially identical to that adopted
by \citet{tak12} and \citet{cho14}. The coordinate system of the model disk
is defined with respect to the central stellar position, which is
derived from the peak position of the ALMA 2.9 mm continuum emission \citep{alm15}.
$\alpha$ and $\delta$ are the positional offsets along the right
ascension and declination, and $x$ and $y$ along the disk minor
and major axes, respectively. The model velocity channel maps
($\equiv S_{\rm model} (\alpha,\delta, v)$) are then expressed as
\begin{equation}
S_{\rm model} (\alpha,\delta, v) = (S_{\rm mom0} (\alpha,\delta) / \sigma \sqrt{2\pi}) \times \exp(\frac{-(v-v_{LOS} (\alpha,\delta))^2}{2.0\sigma^2}),
\end{equation}
\begin{equation}
v_{LOS} (\alpha,\delta) = v_{\rm sys} + \sin(i) v_{\rm kep} (r) \cos(\Phi-\theta),
\end{equation}
\begin{equation}
r = \sqrt{(\frac{x}{\cos i})^2 + y^2},
\end{equation}
\begin{equation}
\alpha = x\cos(\theta) + y\sin(\theta),
\end{equation}
\begin{equation}
\delta = -x\sin(\theta) + y\cos(\theta).
\end{equation}
In the above expressions, $S_{\rm mom0} (\alpha,\delta)$ denotes the moment 0 map
of the model disk, $\sigma$ the internal velocity dispersion, and
$v_{LOS} (\alpha,\delta)$ and $v_{\rm sys}$ are the line-of-sight and
systemic velocities, respectively.
$\theta$ and $\Phi$ are the position angle of the disk major axis
and the azimuthal angle
from the major axis on the disk plane.

The velocity ranges of the fitting are restricted to $V_{LSR}$ = 2.5 - 4.0 km $s^{-1}$ and
10.0 - 11.5 km s$^{-1}$, to extract the velocity components of 
the disk solely. $S_{\rm mom0} (\alpha,\delta)$ are assumed to be the same as the observed
moment 0 maps in the relevant velocity ranges. 
$\sigma$ is adopted to be 0.4 km s$^{-1}$, which is the sound speed at a temperature of 20--30 K. The dynamical
center of the model disk is fixed to be the peak position of the 2.9 mm
continuum emission, and $v_{\rm sys}$ is set to be 7 km s$^{-1}$. The fitting
parameters are thus $M_{\star}$, $\theta$, and $i$, and the $\chi^2$ value
with a given set of the fitting parameters is calculated as
\begin{equation}
\chi^2 = \sum_{\alpha,\delta,v} \left( \frac{S_{\rm obs} (\alpha,\delta, v) -S_{\rm model}^{M_{\star},\theta,i} (\alpha,\delta, v)}{\sigma_{\rm rms}}\right) ^2 / \sum_{\alpha,\delta,v}
\end{equation}
where $S_{\rm obs} (\alpha,\delta, v)$ and $\sigma_{rms}$ denote the observed velocity channel maps and the rms noise level, respectively.

\begin{figure}
\epsscale{1} 
\plotone{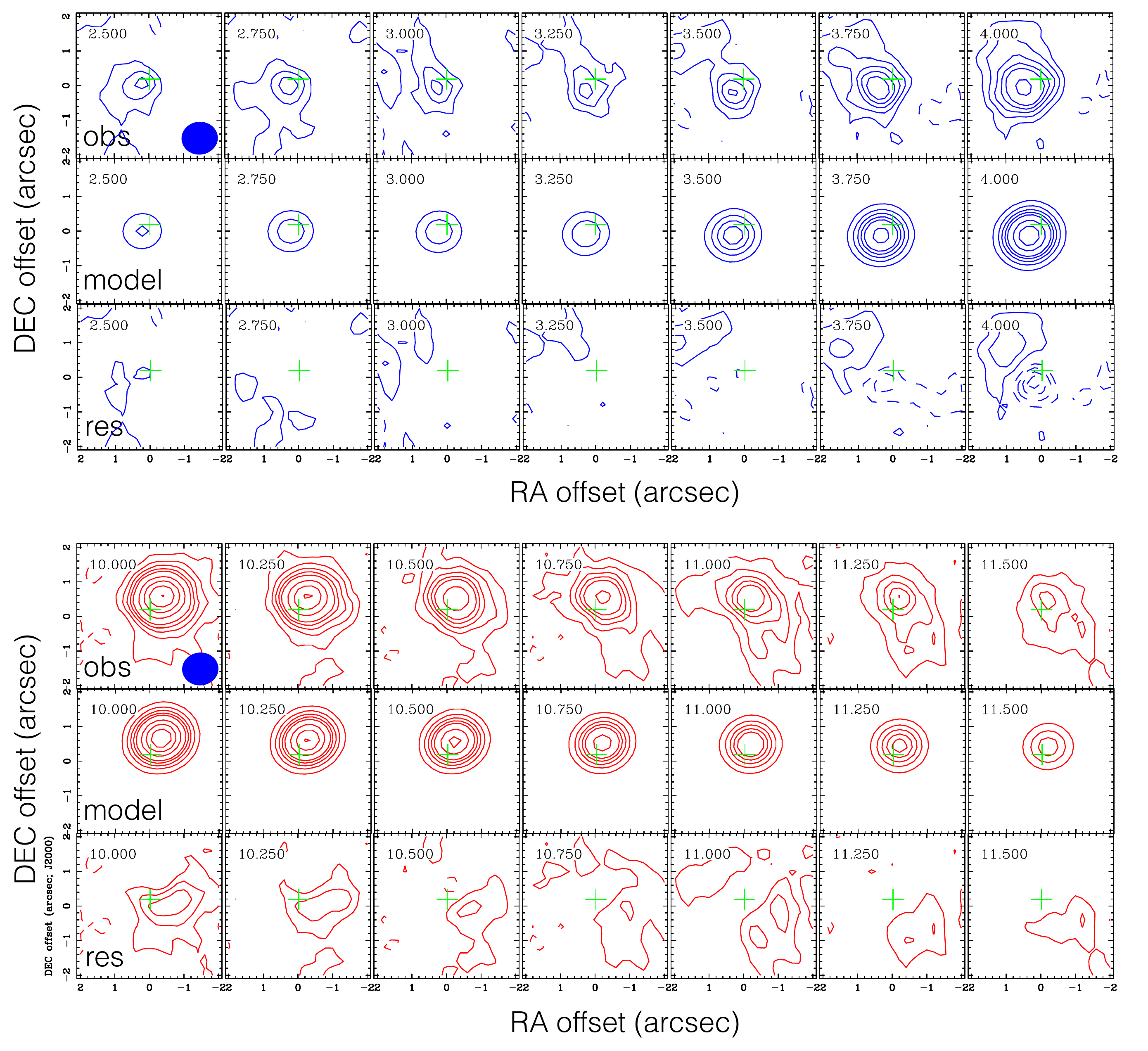}
\caption{Comparison of the observed, model, and the residual velocity channel maps of the  HCO$^+$ (1--0) emission in HL Tau at the highly blueshifted (blue contours) and redshifted velocities (red). Contour levels are -10$\sigma$, -8$\sigma$, -6$\sigma$, -4$\sigma$, -2$\sigma$, 2$\sigma$, 4$\sigma$, 6$\sigma$,8$\sigma$, 10$\sigma$, 12$\sigma$, 14$\sigma$, 16$\sigma$, 18$\sigma$, 20$\sigma$ (1$\sigma$ = 2.4 mJy beam$^{-1}$). Crosses and blue ellipses denote the protostellar position and the synthesized beam (Table \ref{table:imaging}), respectively.}
\label{Fig:kepfit}
\end{figure}

The best-fit parameters are $M_{\star}$ = 1.4 $M_{\odot}$,
$\theta$ = 142$\degr$, and $i$ = 53$\degr$, providing
the minimum $\chi^2$ value of 2.73.
The best-fit $\theta$ and $i$ are consistent with those estimated
from the 2.9 mm continuum emission. 
The derived stellar mass is also consistent with the estimates by \citet{alm15}
and \citet{yen17b}.
Figure \ref{Fig:kepfit} shows the observed (top panels), best-fit model, and the residual
velocity channel maps of the HCO$^{+}$ (1--0) emission. At the higher velocities of $|\Delta V| > 3.5$ km s$^{-1}$, the model
Keplerian disk reproduces the observed velocity channel maps, but at the lower velocities of $|\Delta V| \sim 3$ km s$^{-1}$,
significant residuals are present. At the blueshifted velocity, the residual
emission is predominantly located to the northeast of the protostar, while at the
redshifted velocity to the southwest. As discussed above, these outer components most
likely trace the infalling envelope or contamination from the associated molecular
outflow outside the Keplerian disk, which cannot be modeled with our geometrically-thin Keplerian disk model.

\subsection{SPARX Modeling of the Physical Conditions of the Disk and Envelope around HL Tau}\label{SPARX}
Based on our best-fit Keplerian disk model, we
constructed more detailed models of the disk plus the flattened infalling envelope
around HL Tau. The volume density and temperature profiles of the Keplerian
disk and envelope are defined in the cylindrical coordinates ($r, z$) as \citet{kwo11,oku16}
\begin{eqnarray}\label{eq_11}
\rho(r,|z|)=\frac{(2-\gamma)M_d}{2\pi r_c ^2\sqrt{2\pi}H(r)}\left( \frac{r}{r_c}\right) ^{-\gamma} \exp\left(-\frac{|z|^2}{2H^2(r)} \right)
\end{eqnarray}
\begin{eqnarray}
T(r)=T_{1{\rm au}}\left( \frac{r}{1~{\rm au}}\right) ^{-0.57}
\end{eqnarray}
In the above expressions, $T_{1{\rm au}}$, $r_c$, and $M_d$ denote the temperature at $r$ = 1 $au$,
the disk radius, and the disk mass, respectively. $\gamma$ expresses the
power-law index of the volume density profile. $H(r)$ denotes the disk scale height at a
radius of $r$, and is expressed as
\begin{eqnarray}
H(r) = \sqrt{\frac{c_{s}^{2}r^{3}}{GM_{\star}}},
\end{eqnarray}
where $c_{s}$ is the sound speed.
In our model, the exponential tapering, which represents
the outer cutoff of the disk, is not included, and the power-law density profile
is extended to the outer envelope continuously.
$\gamma$ is set to be $\gamma = 1$
in the disk and $\gamma = 0.5$ in the envelope \citep{oku16}, and $r_c = 100$ au
\citep{alm15}. $\gamma$ of 1.0 and 1.5 in the envelope have also been tested, but the models with $\gamma$ of 0.5 best reproduce the intensity profile of the envelope.
Gas motions in the disk and the envelope in our model are assumed to be only dependent on the radius. 
In the disk, the Keplerian rotation of the central mass of 1.4 $M_{\odot}$ is adopted. 
In the envelope, the free-fall motion with the central mass of 1.4 $M_{\odot}$ and the rotation with a conserved specific angular momentum of 0.078 km $s^{-1}$ pc, which is computed at the disk outer radius, are adopted.

With these physical and dynamical formulae of the disk and envelope and the parameters given in Table \ref{table:sparx},
we performed radiative transfer calculations of the $^{13}$CO (2--1), C$^{18}$O (2--1),
and 1.3 mm dust-continuum
emission with the SPARX (Simulation Platform for
Astrophysical Radiative Xfer) code. 
Because the ratio of the observed peak integrated intensities of the $^{13}$CO (2--1) and (1--0) emission, 16.9 and 3.1 Jy km s$^{-1}$ beam$^{-1}$, is consistent with the expectation from the LTE condition within the uncertainty \citep[this work;][]{hay93}, the LTE condition is assumed in our radiative transfer calculations. 
In addition, we assume that there is no spatial segregation between the gas and dust components, 
and that gas and dust are thermally well coupled.
We computed models with 0.0001 $M_{\odot}$$ < M_d < $0.35  $M_{\odot}$ and 200 K $< T_{1{\rm au}} <$ 650 K for the $\chi^2$ fitting,
and the ranges and steps of the parameters are progressively narrowed down to approach the best fit. 
The image cubes generated with the radiative transfer calculations of the molecular lines are sampled on the observed $uv$ grids,
and the model visibilities are continuum-subtracted, Fourier-transformed, and CLEANed with the same process as that adopted for the observational data.
Then we compare the inner 15$\arcsec$$\times$15$\arcsec$ of the moment 0 maps and the intensity profiles along the major axis at r$<$6$\arcsec$ of the $^{13}$CO (2--1) and C$^{18}$O (2--1) emission between our models and observations, and compute $\chi^2$.

\begin{figure}
\epsscale{1} 
\plotone{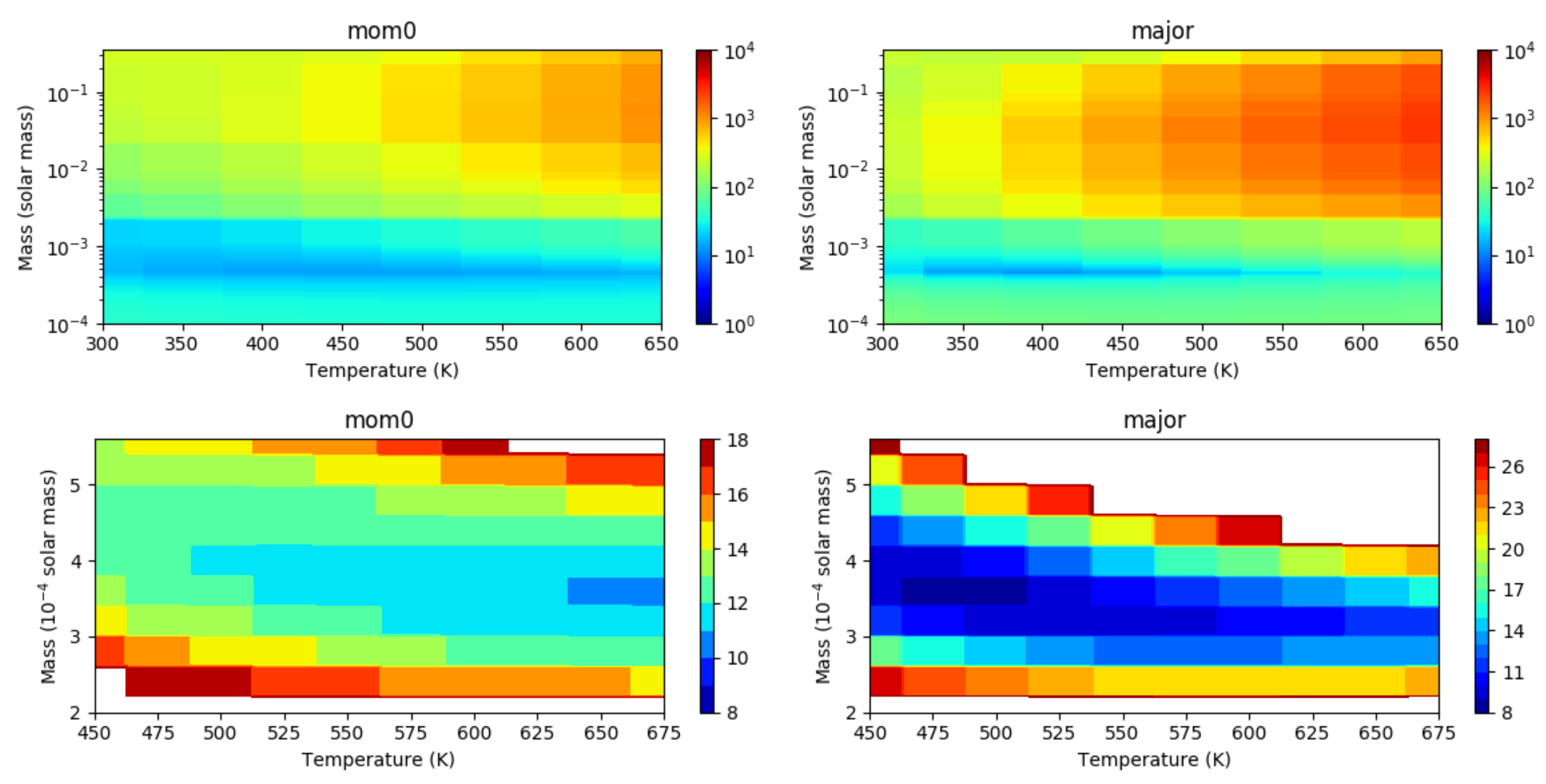}
\caption{$\chi^{2}$ distributions as a function of $M_d$ and $T_{1{\rm au}}$ in our SPARX model fitting to the observed moment 0 maps (left panels) and to the intensity profiles along the major axis (right panels) of the $^{13}$CO (2--1) and C$^{18}$O (2--1) emission. Bottom panels are the zoom-in of the top panels.}
\label{Fig:chi}
\end{figure}

\begin{figure}
\epsscale{1} 
\plotone{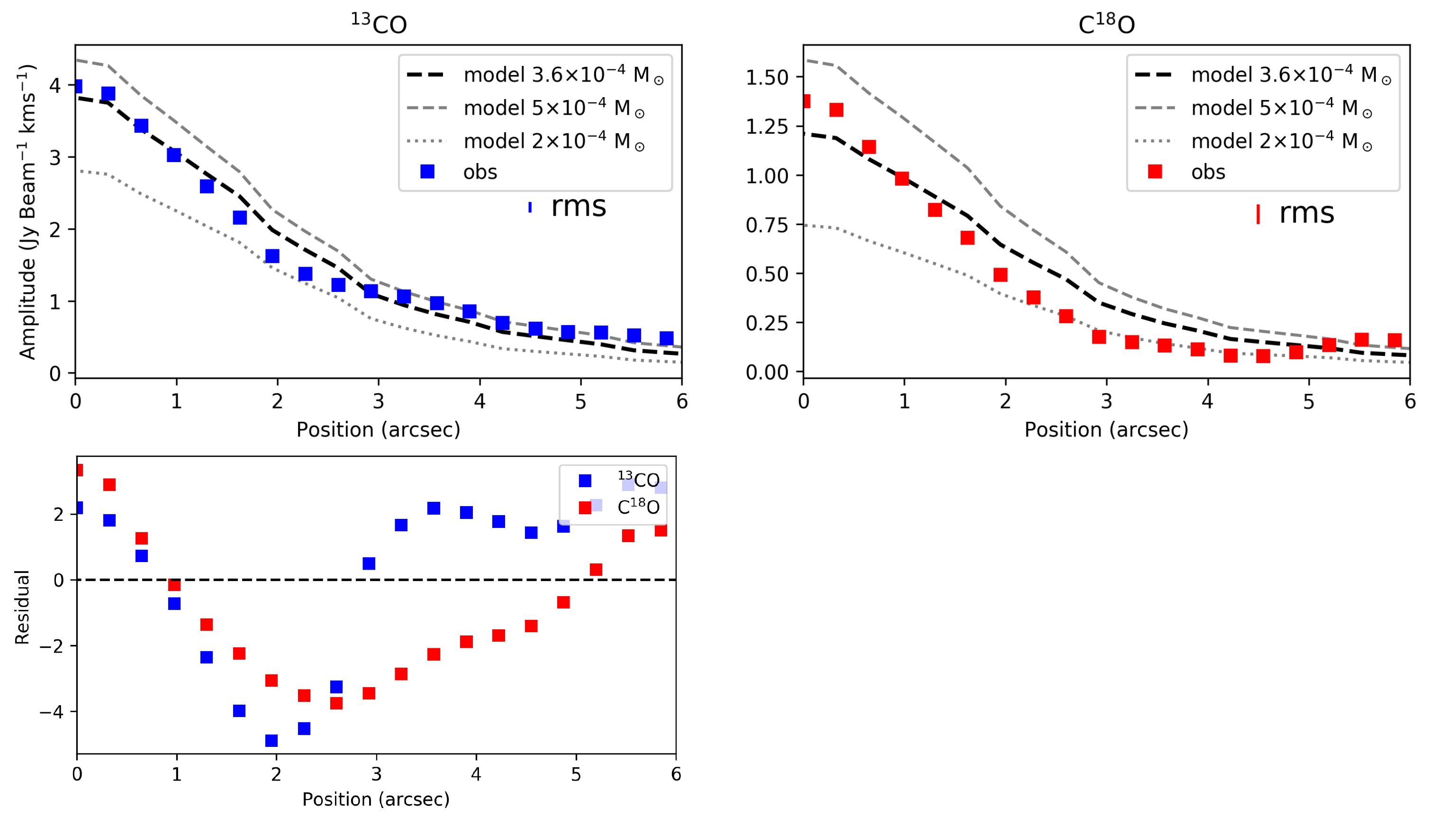}
\caption{Observed (squares) and model (lines) intensity profiles along the major axis of the $^{13}$CO (2--1; top left panel) and C$^{18}$O (2--1; top right panel) emission and the residuals of the $^{13}$CO (blue squares) and C$^{18}$O (red squares) emission after subtracting the best-fit model from the observations in units of the observational noise levels  (bottom panel). In the top panels, black dashed, gray dashed, and gray dotted lines present the intensity profiles from the best-fit model and the models with the estimated upper and lower limits of the disk mass, respectively. The disk masses in these three models are labelled at the upper right corner. The best-fit parameters are $M_d=3.6\times10^{-4}\:M_{\odot}$ and  $T_{1au}=500\:K$. Most residuals are less than 3 $\sigma$. The maximum residual is 5 $\sigma$ at a radius of 2$\arcsec$ in the $^{13}$CO emission.
}
\label{Fig:maj}
\end{figure}

\begin{figure}
\epsscale{1} 
\plotone{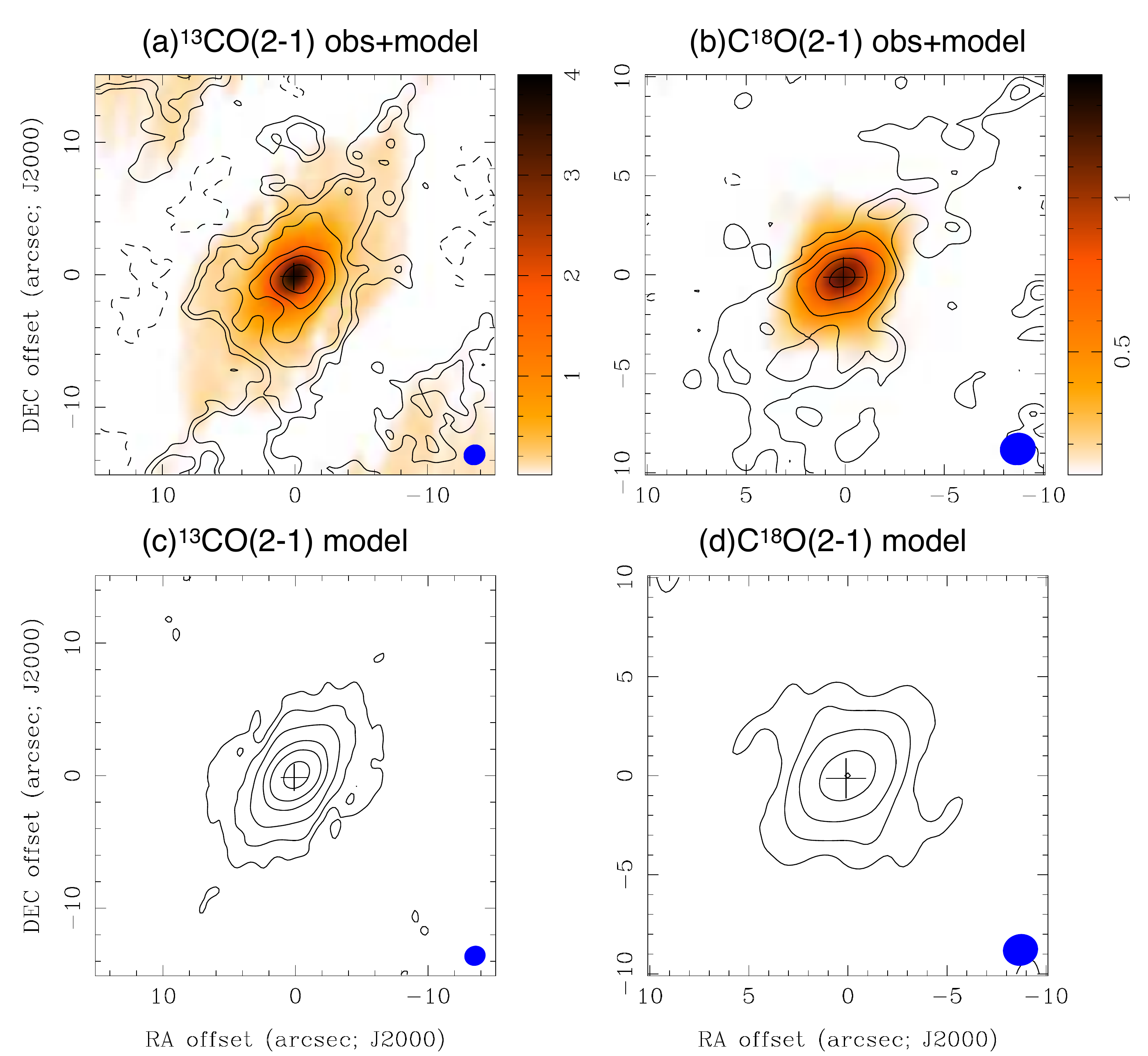}
\caption{Comparisons of the observed (contours) and model (color scale) moment 0 maps of the $^{13}$CO (2--1; upper left panel) and C$^{18}$O (2--1; upper right panel) emission. The model moment 0 maps are obtained from our best-fit model to the intensity profiles along the major axis and are also shown in contours in bottom panels. All the contour levels are the same as those in Figure \ref{Fig:moment0}.}
\label{Fig:sparx13coc18o}
\end{figure}

We first adopted a dust opacity per H$_2$ mass of 8.99 $\times$ 10$^{-3}$ cm$^2$ g$^{-1}$ in our model calculations, which is equivalent to a typical dust-only mass opacity of 0.899 cm$^2$ g$^{-1}$ with a canoncial gas-to-dust mass ratio of 100. 
The fitting results with this dust mass opacity are summarized in Figures \ref{Fig:chi}, \ref{Fig:maj}, and
\ref{Fig:sparx13coc18o}. Figure \ref{Fig:chi}
left and right panels show $\chi^2$ distributions of the fittings to the moment 0 maps and to the intensity profiles
along the major axis, respectively. The best-fit parameters to the moment 0 maps
are $M_d$ = 3.6 $\times$ 10$^{-4}$ $M_{\odot}$ and $T_{1{\rm au}}$ = 675 K,
while those to the intensity profiles are $M_d$ = 3.6 $\times$ 10$^{-4}$ $M_{\odot}$
and $T_{1{\rm au}}$ = 500 K. The minimum reduced $\chi^2$ of the fitting to the moment 0 maps and intensity profiles are 10.8 and 8.7, respectively.
Figure \ref{Fig:maj}
shows the best-fit intensity profiles of the $^{13}$CO and C$^{18}$O emission
and the residuals. The moment 0 maps
of the model and the observations are also shown in Figure \ref{Fig:sparx13coc18o}.
There are possible systematic residuals in our fitting to the $^{13}$CO and C$^{18}$O intensity profiles (lower left panel in Figure \ref{Fig:maj}). We found that the observed intensity profiles cannot be fully reproduced with our models having a simple power-law density profile. Thus, here we aim to provide the upper and lower limits of the disk mass derived from the CO isotopologue lines rather than introducing new parameters in our models to fit the intensity profiles. To estimate the upper and lower limits of the disk mass, we first fixed $T_{1{\rm au}}$ and the envelope to disk mass ratio in our models at the best-fit values, 500 K and 8.5. Then we increased and decreased the disk and envelope masses in the models and generated a series of intensity profiles until all the observed data points were bounded by the model intensity profiles (gray dashed and dotted lines in Figure \ref{Fig:maj}). We adopt that range of the disk mass, 2--5 $\times$ 10$^{-4}$ $M_{\odot}$, as the lower and upper limits. We have also tested the dependence of our estimate on $T_{1{\rm au}}$. We found that the observed $^{13}$CO and C$^{18}$O intensity ratio cannot be reproduced with the models having different $T_{1{\rm au}}$. For example, in the models with $T_{1{\rm au}}$ of 300 K and 700 K, the intensity ratios of $^{13}$CO and C$^{18}$O at the peak are $\sim$2 and $\sim$4, respectively, different from the observed ratio of $\sim$3. 

In addition, we found that the adopted dust mass opacity in our models yields a significantly weaker dust-continuum emission compared to the observations, and that the dust-continuum emission in the models is optically thin.
On the other hand, the observed dust-continuum emission of HL Tau shows a spectral index $\alpha$ of $\sim$2
in the bright rings in the disk. This indicates that the disk emission is partially optically
thick \citep{alm15}, which is expected to suppress the intensity of the molecular-line emission \citep[e.g.,][]{har18}.
Therefore, our estimated gas mass with this dust mass opacity is considered as a lower limit. 

\begin{figure}
\epsscale{1} 
\plotone{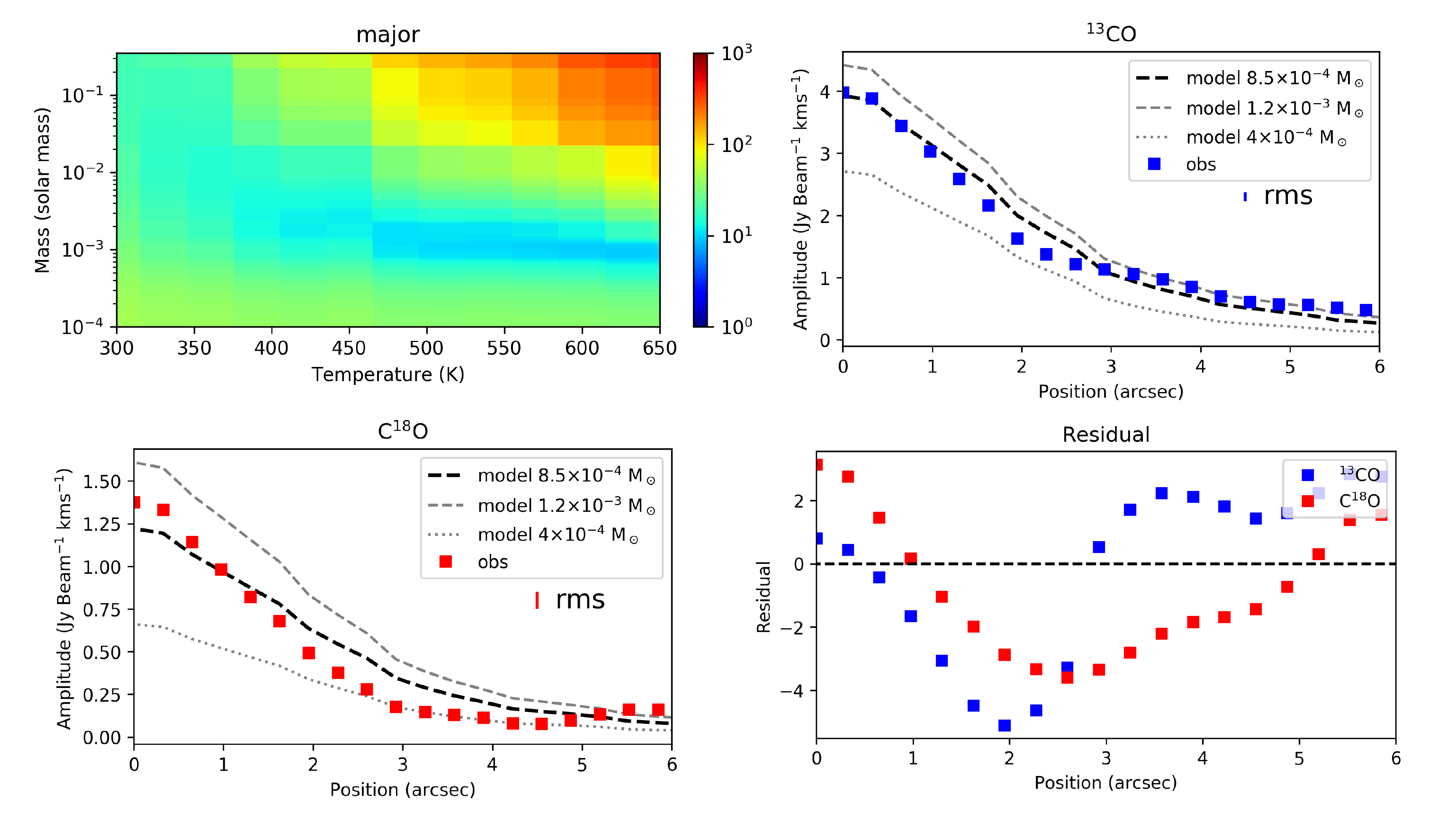}
\caption{Results of the SPARX model fitting to the observed intensity profiles along the major axis of the $^{13}$CO (2--1) and C$^{18}$O (2--1) emission, whose the dust mass opacity is adjusted to have the continuum intensity in the model consistent with the observations. Top left panel shows the $\chi^{2}$ distribution as a function of $M_d$ and $T_{1{\rm au}}$. Top right panel shows the observed (squares) and model (lines) intensity profiles along the major axis of the $^{13}$CO (2--1) emission. Bottom left panel is the same as the top right panel but for the C$^{18}$O (2--1) emission. In the top panels, black dashed, gray dashed, and gray dotted lines present the intensity profiles from the best-fit model and the models with the estimated upper and lower limits of the disk mass, respectively. The disk masses in these three models are labelled at the upper right corner. Bottom right panel shows the residuals of the $^{13}$CO (blue squares) and C$^{18}$O (red squares) emission after subtracting the best-fit model from the observations in units of the observational noise levels.}
\label{Fig:1_times}
\end{figure}

To reproduce the observed continuum intensity, 
for a given set of $M_d$ and $T_{1{\rm au}}$, we first search for a dust mass opacity such that the intensities of the dust-continuum emission in the model and observations are consistent. 
Then, we adopt that dust mass opacity to generate molecular-line images of the model, and perform the $\chi^2$ fitting with the same procedure described above. 
Because in the observations the continuum emission in the envelope is optically thin and does not affect the continuum-subtracted line intensity, 
we only adjust the dust mass opacity in the disk but not in the envelope in our model calculations, 
and the envelope mass and temperature distributions are fixed to be the best-fit parameters obtained from the previous fitting. 
We then vary the disk mass and temperature to search for the best fit.
The resultant $\chi^2$ distribution and the best-fit intensity profiles are 
shown in Figure \ref{Fig:1_times}. The best-fit parameters of the disk
are $M_d$ = 8.5 $\times$ 10$^{-4}$ $M_{\odot}$ and $T_{1{\rm au}}$ = 630 K with a dust opacity per H$_2$ mass of 1.07 cm$^2$ g$^{-1}$.
The minimum reduced $\chi^2$ of the fitting to the intensity profile is 9.5.
Then we followed the same process described above to estimate the upper and lower limits of the disk mass. The range of the disk mass from the CO isotopologue lines is estimated to be 0.4--1.2 $\times$ 10$^{-3}$ $M_{\odot}$ when the dust opacity was adjusted to match the continuum intensities in the models and the observations.
We note that in the HL Tau disk, the dust is likely settled to the mid plane with respect to the gas \citep{kwo11,pin16}, 
and the gas and the dust could have different temperature profiles \citep{oku16, yen16}. 
In our models, there is no spatial segregation between gas and dust, and the gas and dust temperatures are assumed to be the same. 
Therefore, our model calculations with the adjusted dust mass opacity likely overestimate the suppression of the line intensity by the continuum opacity,  
and the disk gas mass estimated from this $\chi^2$ fitting can be considered as an upper limit. 

We note that our best-fit $T_{1{\rm au}}$ of $\sim$ 500 -- 675 K is higher than the value of 310 K derived from the intensity profile of the dust-continuum emission by \citet{oku16}.  
This difference could suggest that the dust-continuum emission traces the mid plane temperature, and our molecular-line intensity likely traces the averaged temperature of the mid plane and surface. 
In addition to the fittings described above, we have also tested the effects of different density profiles and abundance ratios between $^{13}$CO and C$^{18}$O on our fitting results. 
The $\chi^2$ fitting with $\gamma$ = 1 in the disk and $\gamma$ = 1 in the envelope is performed. 
With this density profile, the $\chi^2$ fitting of the intensity profiles yields a higher minimum $\chi^2$ of 10, and the best-fit model exhibits less emission compared to the observation at larger radii.
An abundance ratio of 10 between $^{13}$CO and C$^{18}$O \citep{bri05,smi15} is also tested, 
but this abundance ratio results in a higher $\chi^2$ value compared to our best-fit model with the abundance ratio of 7.3.
We also constructed another model with the C$^{18}$O abundance that is three times lower in the disk and remains unchanged in the envelope to mimic the possible effect of the selective photodissociation in protoplanetary disks, as in \citet{wil14}.
The best-fit parameters become $M_d$ = 3 $\times$ 10$^{-3}$ $M_{\odot}$ and $T_{1{\rm au}}$ = 480 K in this model. The minimum reduced $\chi^2$ of the fitting to the intensity profile is 9.6. The range of the disk mass derived from the CO isotopologue lines is 1.5--4 $\times$ 10$^{-3}$ $M_{\odot}$.

Therefore, considering all the uncertainties in the continuum opacity, and the abundance ratio between $^{13}$CO and C$^{18}$O, and the disk temperature, 
our fitting results suggest that the disk gas mass in HL Tau is most likely in the range of 2--40 $\times$ 10$^{-4}$ $M_{\odot}$. 
On the other hand, the gas mass in the envelope is estimated to be 2.9 $\times$ 10$^{-3}$ $M_{\odot}$ from our best-fit model by integrating the mass in the radial range of 100 -- 1800 au.
Then, the mass infalling rate in the envelope ($\equiv \dot{M}$) can be calculated as, 
\begin{equation}
\dot{M} = 2\pi r \Sigma(r) v_{\rm freefall}(r),
\end{equation}
where $\Sigma(r)$ and $v_{\rm freefall}(r)$ denote the surface density and
the infall velocity at radius $r$, respectively.
$\dot{M}$ is estimated to be 2.5 $\times$ 10$^{-6}$ $M_{\odot}$ $yr^{-1}$ at $r$ = 100 au, 
assuming that the infalling velocity equals the free-fall velocity
toward the central stellar mass of 1.4 $M_{\odot}$, i.e., $v_{\rm freefall}$ ($r$ = 100 au) = 3.4 km s$^{-1}$.

\begin{deluxetable}{lccc}
\tabletypesize{\footnotesize}
\tablecolumns{3}
\tablewidth{0pt}
\tablecaption{SPARX Model Parameters \label{table:sparx}}
\tablehead{Parameter &Disk &Envelope}
\startdata
Simulation Package&\multicolumn{2}{c}{SPARX}\\
Radius (au)&14.25-100&100-1800\\
$M_{\star}$&\multicolumn{2}{c}{1.4 $M_{\odot}$}\\
Position Angle&\multicolumn{2}{c}{142$\degr$}\\
Inclination&\multicolumn{2}{c}{53$\degr$}\\
$^{13}$CO abundance$^{(1), (2)}$&\multicolumn{2}{c}{3.5$\times$10$^{-6}$}\\
C$^{18}$O abundance$^{(1), (2)}$&\multicolumn{2}{c}{4.8$\times$10$^{-7}$}\\
$\gamma$& 1 & 0.5\\
$r_c$& \multicolumn{2}{c}{100 au}\\
\enddata
\tablerefs{(1) \citet{lac94}, (2) \citet{wil94}}
\end{deluxetable}

\section{Discussion}\label{discuss}

\subsection{Gas to Dust Ratio in HL Tau}\label{dtg}
With our SPARX fitting (Section \ref{SPARX}), the gas mass of the HL Tau disk is estimated to be $2-40\times10^{-4}$ $M_{\odot}$ from the CO isotopologue lines.
This estimated disk gas mass is comparable to or even lower than the disk dust mass of $1-3 \times 10^{-3}$ $M_{\odot}$ estimated from the continuum emission \citep{kwo11,car16}, leading to a gas-to-dust mass ratio of 0.07--4 in the disk. 
The gas mass of the envelope is estimated to be $2.9\times10^{-3}$ $M_{\odot}$ with our SPARX fitting of the $^{13}$CO and C$^{18}$O emission, and the envelope dust mass to be $1.2 \times 10^{-3}$ $M_{\odot}$ from the continuum emission, suggesting a gas-to-dust mass ratio of 2.2--2.6 in the envelope.

We note that the gas-to-dust mass ratio in the envelope is underestimated with the SMA observations because the molecular-line emission in the envelope has a higher missing flux of $>80\%$ compared to the continuum emission which has almost no missing flux.  
Thus, we have also estimated gas and dust masses of the envelope with the single-dish measurements. 
The 1.3 mm flux of the continuum emission within the radius of 30$\arcsec$ (4200 au) is 1200 mJy observed with the MPIfR bolometer arrays on the IRAM 30 m telescope \citep{mot01}. 
Our IRAM 30 m observations in the C$^{18}$O (2--1) emission of HL Tau  (see Figure \ref{Fig:spectral}) show that the C$^{18}$O flux within the radius of 30$\arcsec$ is 282 Jy km s$^{-1}$. 
We first subtracted the contribution from the disk to the total fluxes measured with the single-dish observations. 
With our SMA observations and SPARX modeling, the 1.3 mm continuum and C$^{18}$O (2--1) fluxes in the disk are measured to be 680 mJy and 4.45 Jy km s$^{-1}$, respectively. 
Thus, the fluxes in the envelope are estimated to be 520 mJy in the continuum and 278 Jy km s$^{-1}$ in the C$^{18}$O emission. 
On the assumption that the dust temperature and the excitation temperature of C$^{18}$O (2--1) are the same in the envelope and are 15 K, which is estimated from the C$^{18}$O (3--2) and (2--1) intensity ratio (Yen et al.~in prep.), 
 the dust mass is estimated to be $3.3 \times 10^{-3}$ $M_{\odot}$ with $\kappa_\nu = 0.899$ cm$^2$ g$^{-1}$, 
 and the gas mass is estimated to be $9.6 \times 10^{-2}$ $M_{\odot}$ with the LTE condition, the optically-thin approximation, and a C$^{18}$O abundance of $4.8\times10^{-7}$. 
With these values, the gas-to-dust mass ratio in the envelope is estimated to be $\sim$29, a factor of ten higher than the estimate with the interferometric measurements. 
If the temperature in the envelope is higher, then the estimated gas-to-dust mass ratio becomes higher. 
For example, when the assumed temperature in the envelope increases from 15 K to 25 K, the gas-to-dust mass ratio becomes a factor of $\sim$2 higher. 
In addition, our estimated gas-to-dust mass ratio is inversely proportional to the assumed C$^{18}$O abundance, and the ISM C$^{18}$O abundance is typically in a range of 1--5 $\times10^{-7}$ \citep{fre87,lac94,wil94}. 
Therefore, considering these uncertainties in the temperature and CO abundance in the envelope, 
the gas-to-dust mass ratio in the envelope around HL Tau is estimated to be in a range of 29--139, consistent with the typical ISM value of 100 within a factor of two to three. 
If we assume that the gas-to-dust mass ratio in the envelope around HL Tau is the ISM value of 100, 
our estimated gas and dust masses suggest a C$^{18}$O abundance of $1.4 \times 10^{-7}$ in the HL Tau region, 
which is consistent with the typical abundance in ISM.  

By contrast, the disk with a diameter of $\sim$2$\arcsec$ around HL Tau unlikely suffers the effects of the missing flux, 
which is the dominant uncertainty in estimating the gas-to-dust mass ratio in the envelope with the interferometric measurements.
Therefore, our results could suggest that the gas-to-dust mass ratio decreases by a factor of ten from the envelope to the disk in HL Tau, or that the CO abundance is a factor of ten lower in the disk than in the envelope, leading to the low estimated gas-to-dust mass ratio from the CO isotopologue lines in the disk.

If the lower gas-to-dust mass ratio estimated from the CO isotopologue lines in the disk than in the envelope is due to actual depletion of gas on the disk scale, 
the gas mass of the disk in HL Tau is comparable to or even lower than Jupiter mass.
This suggests that the current HL Tau disk is hard to form Jupiter-mass planets, and/or the planet-forming process in HL Tau is already in the final stage, 
if the disk does not further accrete substantial mass from its surrounding envelope. 
Gas in protoplanetary disks can be dissipated by photoevaporation \citep{ale14}, layered accretion \citep{gam96}, and disk winds \citep{bai16,gre15}, 
which possibly decrease gas-to-dust mass ratios in disks, as discussed in \citet{ans16}. 
However, the gas depletion rates predicted in these mechanisms are on the order of $10^{-10}-10^{-8}$ $M_{\odot}$ yr$^{-1}$, 
which is two orders of magnitude lower than the mass infalling rate from the envelope onto the disk in HL Tau ($2.5 \times 10^{-6}$ $M_{\odot}$ yr$^{-1}$). 
Therefore, the low gas-to-dust mass ratio estimated from the CO isotopologue lines in the HL Tau disk is unlikely due to actual depletion of gas because of the ongoing mass accretion from the envelope onto the disk, which replenishes the gas content in the disk.

Low gas-to-dust mass ratios estimated from CO isotopologue lines have often been found in protoplanetary disks around Class \RN{2} young stellar objects. 
One possible explanation is that the disk gas mass derived from CO isotopologue lines is underestimated because the CO abundance in disks is lower than the expectation \citep[e.g.,][]{ans16,lon17,mio17}. 
There are several mechanisms that could lead CO to deplete. 
CO is expected to freeze out onto dust gains in the mid plane of disks where the temperature is lower than 20--25 K \citep{aik01,aik03,aik05,pon08}. 
CO depletion could also occur in protoplanetary disks even when the temperature is higher than the freeze-out temperature because CO can be converted to the other molecules through chemical reactions, which is discussed as the sink mechanism under disk physical conditions in \citet{fur14}. 
In the sink mechanism, CO in the gas phase interacts with helium ion to produce carbon ion, which can react with other carbon-bearing species and produce carbon-chain molecules, and thus, the CO abundance decreases. 
Indeed, the HD (1--0) observations toward a few Class \RN{2} disks show that the CO abundance can be lower than the conventional expectation by one to two orders of magnitude \citep{mcc16,sch16}.
This chemical process to deplete CO is also discussed in \citet{ber14}, \citet{yu16}, and \citet{mio17}.
The time scale of the CO depletion via sink mechanism is around 10$^5$ to 10$^6$ years, which is comparable to the age of HL Tau. 
Therefore, in HL Tau, the estimated gas-to-dust mass ratio in the disk that is a factor of ten lower than in the envelope could be due to the low CO abundance in the disk caused by the additional chemical process occurred in the disk but not in the envelope, 
and the disk gas mass can be underestimated by one to two orders of magnitude.

\subsection{Comparison with Class \RN{1} and Class \RN{2} sources}

The SED classification suggests that HL Tau is in the evolutionary stage from Class \RN{1} to Class \RN{2} \citep{ken95,rob07}.  
The disk dust mass of $1\mbox{--}3 \times 10^{-3}$ $M_\odot$ and the envelope-to-disk mass ratio of $\sim$0.4--1.2 in HL Tau are comparable to other Class \RN{1} protostars observed with CARMA in the 1.3 mm continuum emission by \citet{she17}, except for the extreme sources IRAS 04181+2654A \& B having very low-mass disks buried in a very massive envelope. 
On the other hand, the disk dust mass in HL Tau is a factor of a few to ten higher than that in protoplanetary disks around Class \RN{2} young stellar objects with similar stellar masses of 1--2 $M_\sun$. 
Thus, considering the high disk dust mass and envelope-to-disk mass ratio, the HL Tau disk is most likely less evolved compared to those Class \RN{2} disks, 
and its physical properties are more similar to those of the Class \RN{1} disks.

There are several Class \RN{1} disks where the disk dust and gas masses have been estimated from the millimeter continuum and the CO isotopologue lines \citep[e.g.,][]{har14, yen14}.
In \citet{har14}, the disk dust mass is estimated by fitting the continuum emission in the $uv$-plane with different disk models,  
and the disk gas mass is estimated from the integrated C$^{18}$O (2--1) flux within the disk area, where the Keplerian rotation is identified, on the assumptions of $\tau=0.5$ and $T_{\rm ex}=40$ K. 
In \citet{yen14}, the disk dust mass is estimated from the continuum flux measured by Gaussian fitting of the visibility amplitude profile. 
The disk gas mass is estimated by fitting the C$^{18}$O (2--1) high velocity emission with Keplerian disk models. 
These results of the Class \RN{1} disks as well as our HL Tau results are summarized in Table \ref{table:different}. 
Since the adopted C$^{18}$O abundance and dust mass opacity ($\kappa_{1.3mm}$) in the disk are different in different works, these values are also listed in the table. 

The estimated gas-to-dust mass ratios from the continuum and CO isotopologue lines in these Class \RN{1} disks are all consistent with the typical ISM value of 100 within a factor of five. 
By contrast, the estimated gas-to-dust mass ratio in the HL Tau disk is significantly lower than that in the other Class I disks by one to two orders of magnitude even after considering the difference in the adopted C$^{18}$O abundances and dust mass opacities. 
As the HL Tau disk is less evolved than Class \RN{2} disks and has physical properties more similar to Class \RN{1} disks, 
the presence of the low gas-to-dust mass ratio of $<$10 estimated from the CO isotopologue lines in the HL Tau disk, which is not commonly seen in other Class \RN{1} disks, suggests that the chemical process causing a significant decrease of the CO abundance in protoplanetary disks could start at the end of the Class \RN{1} stage. 
Nevertheless, the sample size of Class \RN{1} disks with both measurements of dust and gas masses from the continuum and CO isotopologue lines is still limited. 
Systematic studies of dust and gas masses of Class \RN{1} disks is crucial to constrain when the chemical process causing the CO depletion occurs and its chemical time scale.

\begin{deluxetable}{lcccccc}
\tablecolumns{6}
\tablewidth{0pt}
\tablecaption{The disk dust and gas masses of different Class \RN{1} sources.\label{table:different}}
\tablehead{
\colhead{source}&\colhead{disk}&\colhead{disk}&\colhead{inferred disk}&\colhead{assumed}&\colhead{$\kappa_{1.3mm}$}\\
\colhead{}&\colhead{dust mass ($M_{\odot}$)}&\colhead{gas mass ($M_{\odot}$)}&\colhead{gas-to-dust mass ratio}&\colhead{C$^{18}$O abundance}&\colhead{in the disk}
}
\startdata
HL Tau & $1-3\times10^{-3}$ & $2-40\times10^{-4}$& $0.07-4.0$&4.8$\times$10$^{-7}$&1.0$^{(2)}$\\
TMC1A$^{(1)}$ & $4.1-4.9\times10^{-4}$  & $7.5\times10^{-2}$ & $153-183$&1$\times$10$^{-7}$&0.83\\
TMC1$^{(1)}$& $4.6-5.4\times10^{-5}$ & $2.4\times10^{-2}$ & $444-522$&1$\times$10$^{-7}$&0.83\\
TMR1$^{(1)}$& $1-1.5\times10^{-4}$  & $1\times10^{-2}$ & $67-100$&1$\times$10$^{-7}$&0.83\\
L1536$^{(1)}$& $1.9-2.4\times10^{-4}$ & $6.8\times10^{-3}$& $28-36$&1$\times$10$^{-7}$&0.83\\
L1489 IRS$^{(2)}$& $3-7\times10^{-5}$ & $0.41-1.8\times10^{-2}$& $59-600$&3$\times$10$^{-7}$&2.3\\
\enddata
\tablerefs{(1) \citet{har14} , (2) \citet{kwo11}, (3) \citet{yen14}}
\end{deluxetable}

\section{Conclusions}\label{sum}
We have conducted SMA observations of the candidate of the planet-forming protostar, HL Tau, in the $^{13}$CO ($J$=2--1), C$^{18}$O ($J$=2--1), SO ($J_N$=5$_6$--4$_5$), and the 1.3 mm dust-continuum emission. We have also reanalyzed the ALMA long baseline data of the HCO$^{+}$ ($J$=1--0) emission. The purpose of our project is to unveil the physical and chemical conditions of the protostellar envelope and the central planet-forming disk. The main results are summarized below.

\begin{enumerate}
\item The 1.3 mm dust-continuum emission taken with the SMA consists of a compact (0$\farcs$8 $\times$ 0$\farcs$5) component and an extended (6$\farcs$5 $\times$ 4$\farcs$3) component. The size, position angle ($\sim$ 140$\degr$), and the flux density (0.68 Jy) of the compact component are consistent with those of the central disk observed
with ALMA and CARMA. Thus, the compact component observed with SMA also traces the disk. The extended component most likely traces the surrounding envelope. The total (disk + envelope) flux density observed with the SMA matches with the total flux density measured with the IRAM 30 m. The estimated dust mass of the envelope is $1.2 \times 10^{-3}$ $M_{\odot}$.

\item The molecular-line emission exhibits both compact ($\sim$ 200 au) and extended ($\sim$ 1000 au) components, except for the SO emission showing a compact component only. The $^{13}$CO, C$^{18}$O, and the HCO$^+$ image cubes can be decomposed into high- ($> \pm$ 3.0 km s$^{-1}$) and low-velocity ($< \pm$ 3.0 km s$^{-1}$) components with distinct velocity structures, which correspond to the compact and extended components, respectively. The high-velocity components trace the central disk, as seen in the dust-continuum emission, and exhibit a velocity gradient along the southeast (blueshifted) to northwest (redshifted) direction, which is the direction of the disk major axis. This velocity gradient observed at the high velocity traces the disk rotation. On the other hand, the low-velocity components show a velocity gradient along the northeast (blueshifted) to southwest (redshifted) direction, tracing the infalling motion in the protostellar envelope and the possible contamination from the outflow.

\item Our $\chi^2$ fitting to the high-velocity HCO$^+$ emission demonstrates that the high-velocity emission can be satisfactorily modeled with the geometrically-thin Keplerian disk with the central stellar mass of 1.4 $M_{\odot}$, and disk position and inclination angles of 142$\degr$ and 53$\degr$, respectively. 

\item Based on our fitting results of the Keplerian disk models, we have constructed more detailed physical models of the disk and envelope around HL Tau, computed radiative transfer with the SPARX code, and performed $\chi^2$ fitting to the observed intensity profiles of the $^{13}$CO and C$^{18}$O emission along the major axis of the disk and the envelope. The difference between our best-fit model and the observations is generally less than 3 $\sigma$ with a peak of 5 $\sigma$ at a radius of 2$\arcsec$. Considering the uncertainties due to the abundance ratio between $^{13}$CO and C$^{18}$O, the disk temperature, and the continuum opacity, which suppresses the molecular-line intensity, we estimate the disk gas mass in HL Tau to be $2\mbox{--}40 \times 10^{-4}$ $M_\sun$ with our modeling of the SMA observations, 
while the envelope gas mass is estimated to be $2.9\times10^{-3}$ $M_\sun$.

\item With the disk dust mass in the literature and our estimated disk gas mass from the CO isotopologue lines, the gas-to-dust mass ratio in the HL Tau disk is estimated to be $<$10. The gas-to-dust mass in the envelope is estimated to be around 100 after correcting for the missing flux and considering the uncertainties in the $^{13}$CO and C$^{18}$O abundances and the temperature in the envelope. In protoplanetary disks, CO could deplete through chemical processes even at temperature higher than its freeze-out temperature, resulting in underestimated gas mass and thus gas-to-dust mass ratio from CO isotopologue lines. On the other hand, in HL Tau, the mass infalling rate from the envelope onto the disk is orders of magnitude higher than the mass dissipation rates theoretically expected in protoplanetary disks. Therefore, the gas-to-dust mass ratio in the HL Tau disk that is a factor of ten lower than in the envelope is likely caused by the CO depletion rather than the depletion of gas in the disk. 

\item The comparison with other Class \RN{1} and \RN{2} sources shows that HL Tau is physically similar to Class \RN{1} sources with a high disk dust mass and a high envelope-to-disk mass ratio, and that the HL Tau disk has a low gas-to-dust mass ratio estimated from CO isotopologue lines as many other Class \RN{2} sources, which is not commonly seen in Class \RN{1} sources. This could suggest that the CO depletion in protoplanetary disks, which results in the low gas-to-dust mass ratio estimated from CO isotopologue lines, occurs at the end of the Class \RN{1} stage. In addition, such the CO depletion is not observed in the protostellar envelope around HL Tau. More observations and analyses on sources at the late Class \RN{1} stage are required to constrain the time when the CO depletion happens in protoplanetary disks.

\end{enumerate}

\acknowledgments
This work was supported by NAOJ ALMA Scientific Research Grant Number 2017-04A. S.T. acknowledges PSPS KAKENHI Grant no. JP16H07086
and JP18K03703 in support of this work. N.H. acknowledges a grant from the Ministry of Science and Technology (MoST) of Taiwan (MosT 107-2119-M-001-029). Y.A. acknowledges a grant from the Ministry of Science and Technology (MoST) of Taiwan (MosT 106-2811-M-001-133). This paper makes use of the following ALMA data: ADS/JAO.ALMA\#2011.0.00015.SV. ALMA is a partnership of ESO (representing its member states), NSF (USA) and NINS (Japan), together with NRC (Canada) and NSC and ASIAA (Taiwan) and KASI (Republic of Korea), in cooperation with the Republic of Chile. The Joint ALMA Observatory is operated by ESO, AUI/NRAO and NAOJ.

\software{CASA \citep{mcm07}, \\MIR IDL (https://www.cfa.harvard.edu/rtdc/SMAdata/process/mir/), MIRIAD \citep{sau95}, \\GILDAS/CLASS (http://www.iram.fr/IRAMFR/GILDAS), SPARX}

\clearpage

\end{document}